\documentclass[fleqn,usenatbib]{mnras}

\usepackage[T1]{fontenc}

\DeclareRobustCommand{\VAN}[3]{#2}
\let\VANthebibliography\thebibliography
\def\thebibliography{\DeclareRobustCommand{\VAN}[3]{##3}\VANthebibliography}

\usepackage{graphicx}	
\usepackage{amsmath}	
\usepackage{amssymb}	
\usepackage{newtxtext,newtxmath}

\title[Search for OB associations in {\sl Gaia} EDR3]{Search for OB associations in {\sl Gaia} early Data Release 3}

\author[A. A. Chemel et al.]{
Alexander A. Chemel,$^{1,2,3,4}$\thanks{E-mail: tchemel.sash@yandex.ru}
Richard de Grijs$^{3,4}$
Elena V. Glushkova,$^{1,2}$ and
Andrey K. Dambis$^{2}$
\\
$^{1}$Faculty of Physics, Lomonosov Moscow State University, 1, bld. 2, Leninskie Gory, Moscow, 119992, Russia\\
$^{2}$Sternberg Astronomical Institute, Lomonosov Moscow State University, 13, Universitetskii prospect, Moscow, 119992, Russia\\
$^{3}$School of Mathematical and Physical Sciences, Macquarie University, Balaclava Road, Sydney, NSW 2109, Australia\\
$^{4}$Research Centre for Astronomy, Astrophysics \& Astrophotonics, Macquarie University, Balaclava Road, Sydney, NSW 2109, Australia\\
}

\date{Accepted XXX. Received YYY; in original form ZZZ}

\pubyear{2022}

\begin{document}
\label{firstpage}
\pagerange{\pageref{firstpage}--\pageref{lastpage}}
\maketitle

\begin{abstract}
The distribution of young stars into OB associations has long been in need of updating. High-precision {\sl Gaia} early Data Release 3 astrometry, coupled with modern machine-learning methods, allows this to be done. We have compiled a well-defined sample which includes OB stars and young open clusters, in total comprising about 47,700 objects. To break the sample down into groupings resembling associations, we applied the HDBSCAN$^{*}$ clustering algorithm. We used a Monte Carlo method to estimate the kinematic ages of the resulting clusters and the Student's $t$-test to assess the significance of the linear correlations between proper motions and coordinates, indicating the presence of possible cluster expansion signatures. The ages of the majority of clusters demonstrating a general expansion at a 1$\sigma$ confidence level are several tens of Myr, which is in agreement with the expected ages of OB associations. We found 32 open clusters which turned out to be members of the resulting groupings; their ages are consistent with one another within the uncertainties. Comparison of the clusters thus obtained with the historical composition of OB associations in the literature shows a correspondence between their positions in the Galaxy but an apparent absence of good one-to-one stellar matches. Therefore, we suggest that the historical composition of OB associations needs to be revised.
\end{abstract}




\section{Introduction}

The concept of OB associations as gravitationally unbound groups composed of young O and B stars was introduced in the middle of the previous century by Victor A. Ambartsumian, along with the concept of T and R associations \citep{Amb1947}, although the existence of loose groups of O- and B-type stars had been known since the first works on the spectral classification of bright stars became available. Following \citet{Amb1947}, studies of OB associations were carried out and published in a number of classical papers in the previous century: (i) \citet{Blaauw1952a, Blaauw1952b} obtained direct confirmation of Ambartsumian's hypothesis of the disconnection and expansion of OB associations and estimated the ages of the II Persei (II Per) and Scorpius--Centaurus (Sco--Cen) associations based on the nature of their expansion; (ii) the first lists of OB associations were published by \citet{Markarian1951} and \citet{Morgan1953}; (iii) \citet{Morgan1952} studied signs of the presence of Galactic spiral structure based on the spatial distributions of O- and B-type stars; and (iv) \citet{Becker1963} presented distances to 156 clusters and associations. \citet{Blaauw1964}, which represents a study of the intrinsic properties of OB associations within 1 kpc from the Sun, was an important step for further research in this area. The {\sl Hipparcos} catalogue, published in 1997, became an important benchmark to study the kinematics and spatial distributions of stars in the solar neighbourhood, including young OB stars, that is, potential members of associations. For example, a comprehensive study was carried out by \citet{Zeeuw1999}, who presented an updated census of association members within 1 kpc from the Sun. The release of the {\sl Gaia} mission results, in turn, expands the horizons for new research and allows one to move on to studies of increasingly distant regions of the Galactic disc.

Here, we deal with an expanded concept of young stellar associations compared with previous definitions, defined as gravitationally unbound groups of young stars of all spectral types. OB stars, hot massive and luminous main-sequence objects, are distinguished simply because they are easily identifiable young-age tracers. This is why researchers tend to refer to young stellar associations as `OB associations'. Studies of objects of this type are important for a number of reasons. First, since the overwhelming majority of stars with masses $> 0.5$ M$_{\sun}$ are formed in groups \citep{Lada2003}, OB associations are among the main object types through which one can trace active star formation in the Galaxy. Second, the origin of these objects is not yet fully understood. Key to solving this problem are studies of the associations' kinematic and spatial structure and substructure \citep{Lim2019}, as well as of the kinematics of their expansion and determination of their ages. Several attempts have been made to distinguish stellar groups among the OB supergiants in the Milky Way's thin disc \citep[e.g.,][]{Humphreys1978, BlahaHump1989, Garmany1992, ME1995}. However, the relatively modest sample sizes (fewer than 5000 stars contained in the largest of these studies), the insufficient accuracy of the astrometric data and, often, the simple algorithms used to identify stellar groups, did not allow for selection of a large number of OB associations with high reliability. This might partially explain why researchers still retain the list of associations as they were identified more than 40 years ago, based on now rather outdated data which was also analysed in a subjective manner. 

This situation changed dramatically after the first results of the {\sl Gaia} mission were made public \citep{GaiaMission2016}, and especially after the second data release, {\sl Gaia} DR2 \citep{GaiaDR2_2018}. By now, the community has acquired significantly more accurate information about parallaxes and proper motions for an unprecedented number of approximately 1.5 billion stars in our Galaxy. This has made it possible to increase the number of objects under consideration and expand the range of heliocentric distances studied. In particular, this made it possible to establish that OB associations are not a product of the disruption of initially bound stellar aggregations as a result of gas swept out by young stars, as was previously thought \citep{Hills1980, Brown1997, Kroupa2001, Goodwin2006, Baumgardt2007}, but they are born unbound and inherit the kinematic and structural properties of their parent giant molecular clouds \citep{Wright2018, Ward2018, Lim2019, Ward2020}. Thus, the density spectrum of stellar groups is continuous \citep{Allen2007, Evans2009, Lamb2010}, and modern OB associations are not necessarily subject to global expansion \citep{MD17, MD20}. 

Some attempts to search for OB associations using {\sl Gaia} DR2 data have already been undertaken. For example, \citet{Ward2020} searched for OB associations within 4 kpc from the Sun. However, clustering analysis was carried out by only taking into account the spatial positions of stars in the heliocentric Cartesian coordinate system, without considering proper-motion space; yet, the proximity of stars in this latter space is also a very important criterion to decide whether or not they belong to a group of common origin. In addition, the velocity dispersions of the resulting associations are too high for groups that were only recently formed from gas with a characteristic sound speed of 10 km~s$^{-1}$. For many groups, the one-dimensional velocity dispersion in the plane of the sky exceeds 10 km~s$^{-1}$ and in some individual cases reaches almost 20 km~s$^{-1}$ \citep[][their Table A1]{Ward2020}. Thus, this area remains of high current interest and provides opportunities for further research. In particular, now is the opportune time to establish a new compilation of young associations in the extended solar neighbourhood. This can be done by revisiting the partitioning of young (primarily OB) tracer stars into groups using modern techniques of clustering analysis and by incorporating extensive astrometric data provided by the {\sl Gaia} mission, combined with extensive spectroscopic data, including those of the Large Sky Area Multi-Object Fibre Spectroscopic Telescope (LAMOST) surveys \citep{Deng2012}.

To search for OB associations among young objects in the Galactic thin disc, we used the the most modern HDBSCAN$^*$ clustering analysis algorithm \citep{hdbscan2017}, which has proved successful for solving data clustering problems in stellar astronomy. This algorithm is in many respects an improved version of the DBSCAN algorithm \citep{Ester96adensity-based}, which has also been successfully used to address numerous astronomical problems over the past 20 years, e.g., in searching for open clusters (OCs). The HDBSCAN$^*$ algorithm is good for solving our specific problem, i.e., to search for OB associations among our sample of young objects, taking into account their characteristic properties, such as the often large sizes of OB associations, in both coordinate and velocity space, as well as their low densities compared with OCs. 

We compiled information from many sources, including LAMOST DR5 \citep{LamostDR5}, \citet{Xu2018} and the \citet{Skiff} catalogue of spectral classifications. The resulting sample is one of the largest used to date in the search for OB associations; it includes more than 47,700 objects. The source of our proper motions and parallaxes is {\sl Gaia} early Data Release 3 \citep[EDR3; ][]{GaiaEDR3_2021} and geometric distances were taken from \citet{BJ2021}. Our sample includes not only OB stars but also young OCs from the up-to-date list of \citet{Dias2021}.

Section~\ref{sample} describes in detail the process used for our final sample selection, as well as the approach taken to clean it from objects characterised by low-quality astrometric data. Section~\ref{algo} describes the algorithm used to search for groups among the OB stars. Finally, Section~\ref{res} summarises the main results and offers prospects and directions for follow-up research.

\section{The Sample}\label{sample}

In order to search for OB associations, we compiled a sample which included not only O- and early B-type stars (with masses $\geq 20$ M$_{\sun}$) but also young OCs.

We started with a list of 72,550 OB stars from the \citet{Skiff} catalogue of spectral classifications, which covers the entire celestial sphere. We supplemented our sample with OB stars from LAMOST DR5 and \citet{Xu2018}, which provided information about an additional 10,482 objects. In addition, the sample was supplemented with 1743 OCs from the \citet{Dias2021} catalogue, based on {\sl Gaia} DR2 data. 

First, from the list of objects obtained this way, those that were located outside the Galactic disc were excluded, i.e., those objects having Galactic latitudes $|b| \geq 20 \degr$ were dismissed. We also retained only 955 young OCs whose ages satisfied the condition $\log t (\mbox{ yr}) \leq 8.3$, since older objects are unlikely members of OB associations with a characteristic age of several tens of Myr. 

To carry out the cross-match procedure with the {\sl Gaia} EDR3 catalogue, we applied coordinate propagation for the epoch J2016.0, as follows: 
\begin{enumerate}
	\item A preliminary cross-match of the entire list of stars with the {\sl Gaia} EDR3 catalogue adopting a deliberately large radius of $5\arcsec$;
	\item Use of high-precision proper motions from {\sl Gaia} EDR3 for linear propagation of all equatorial coordinates to the epoch J2016.0; and
	\item Exclusion of all stars which, taking into account the updated coordinates, did not fall into a circular area with a radius of $1.\arcsec5$ around the positions from {\sl Gaia} EDR3. This is equivalent to a $1.\arcsec5$ radius cross-match using J2016.0 equatorial coordinates.
\end{enumerate}
We also used {\sl Gaia} EDR3 \textit{source\_id} to perform a cross-match with the \citet{BJ2021} catalogue. The latter served as a source of information for the geometric distances to the sample stars and their uncertainties.

The next step was to clean the sample from stars with low-quality astrometric data. We excluded all stars with $N_{\mathrm{per}}$ (representing the number of visibility periods used in the astrometric solution; {\sl Gaia} EDR3) $< 8$ and RUWE (renormalised unit weight error; {\sl Gaia} EDR3) $> 1.25$ \citep[see][]{Penoyre2021}. Following application of these quality criteria, we obtained a list containing 50,777 objects. We note that 96 per cent of the remaining stars have a relative {\sl Gaia} EDR3 parallax uncertainty better than 0.2, so no additional criteria involving parallax uncertainties were applied.

\begin{figure}
	\includegraphics[width=\columnwidth]{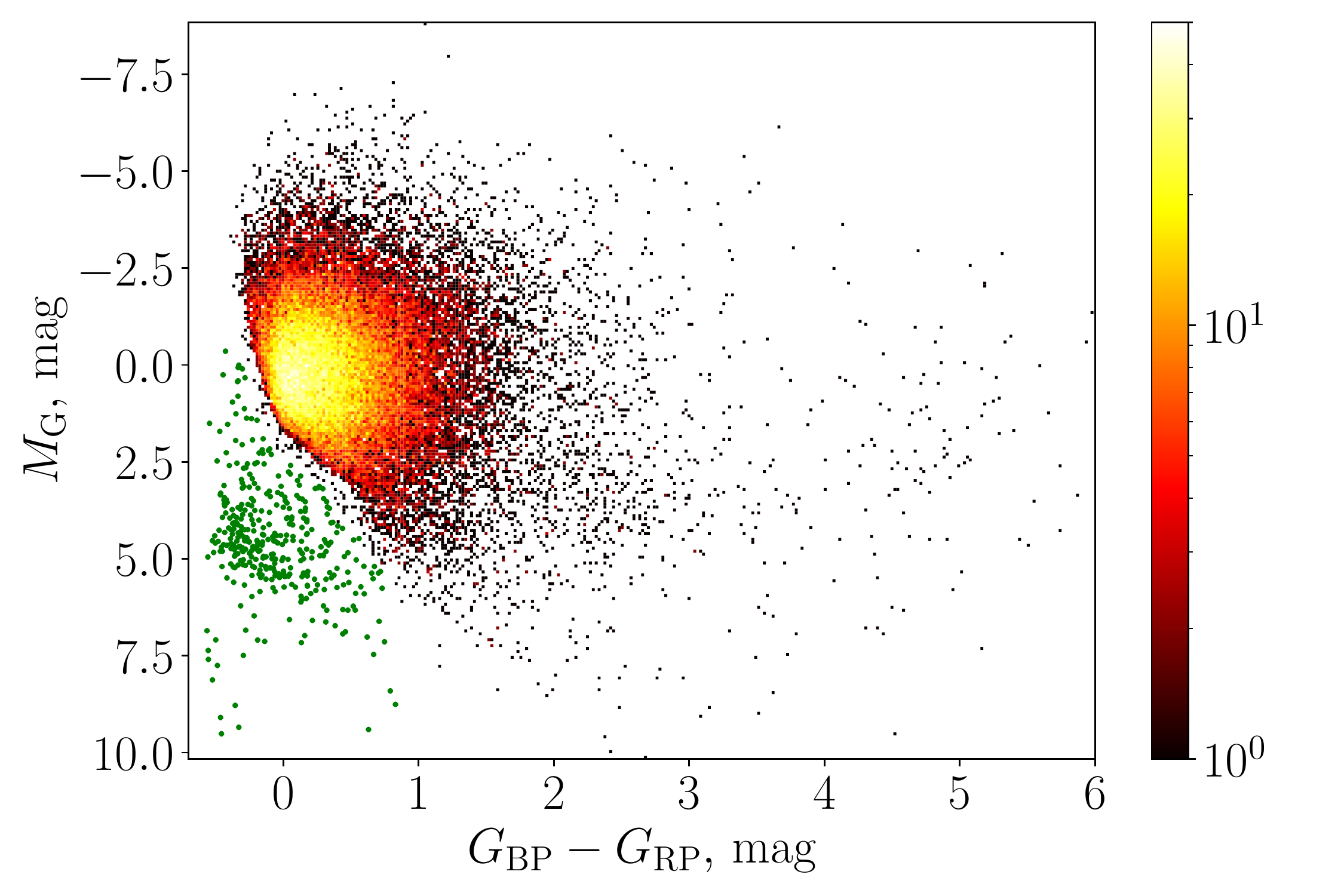}
	\caption{Hertzsprung--Russell diagram for sample stars, plotted in $(G_{\mathrm{BP}} - G_{\mathrm{RP}})$ colour--absolute $M_{\mathrm G}$ magnitude parameter space. The latter was calculated based on the geometric distances from \citet{BJ2021}, without correcting for interstellar reddening. The green dots correspond to sub-main-sequence stars which we have ruled out given their potential nature as white dwarfs or central stars of planetary nebulae. The colour scale for main sequence stars corresponds to the logarithm of their density in the CMD.}
	\label{fig:cmd_removed}
\end{figure}

Since old objects that cannot be present in stellar associations, namely, white dwarfs and central stars of planetary nebulae, also fall into the region of spectral types O and B, we reduced such contamination of our sample by excluding stellar objects located below the main sequence in the Hertzsprung--Russell diagram. The stars we removed are shown in Figure~\ref{fig:cmd_removed} using green dots. The total number of stars thus excluded was 379, or approximately 0.7 per cent of the sample.

Finally, we excluded 3397 stars which were likely members of OCs already contained in the sample. To achieve this, we used a list of members of OCs from \citet{CG2018}, retaining those with a membership probability of 50 per cent or higher. The need to apply this step is owing to the fact that we are not interested in triggering our algorithm on small and relatively dense formations like OCs. Instead, we treat OCs as single data points in the input data.

\begin{figure}
	\includegraphics[width=\columnwidth]{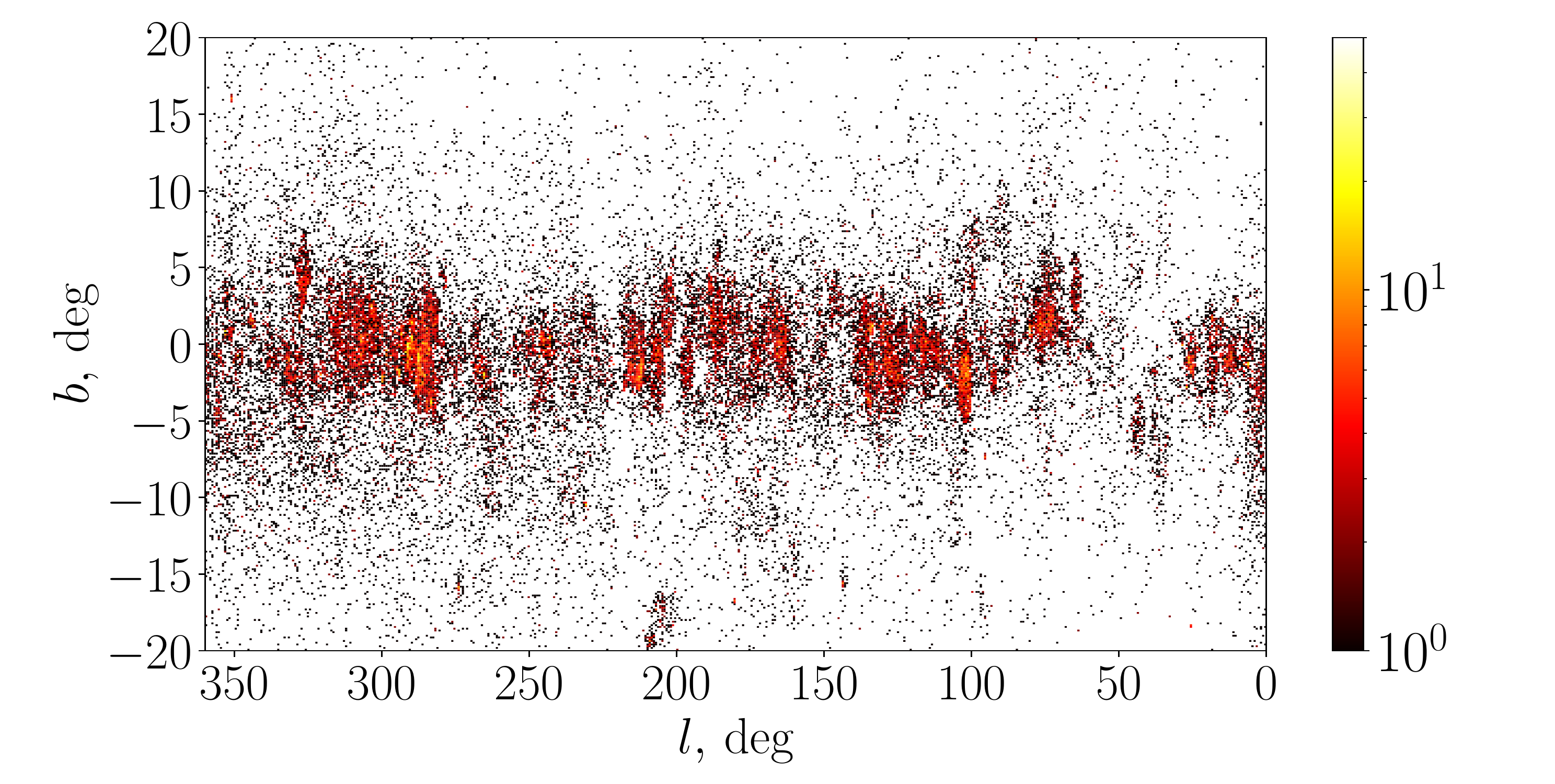}
	\caption{Distribution of our sample objects on the sky in Galactic coordinates. The colour scale corresponds to the logarithm of their density.}
	\label{fig:sample_sky}
\end{figure}

\begin{figure}
	\includegraphics[width=\columnwidth]{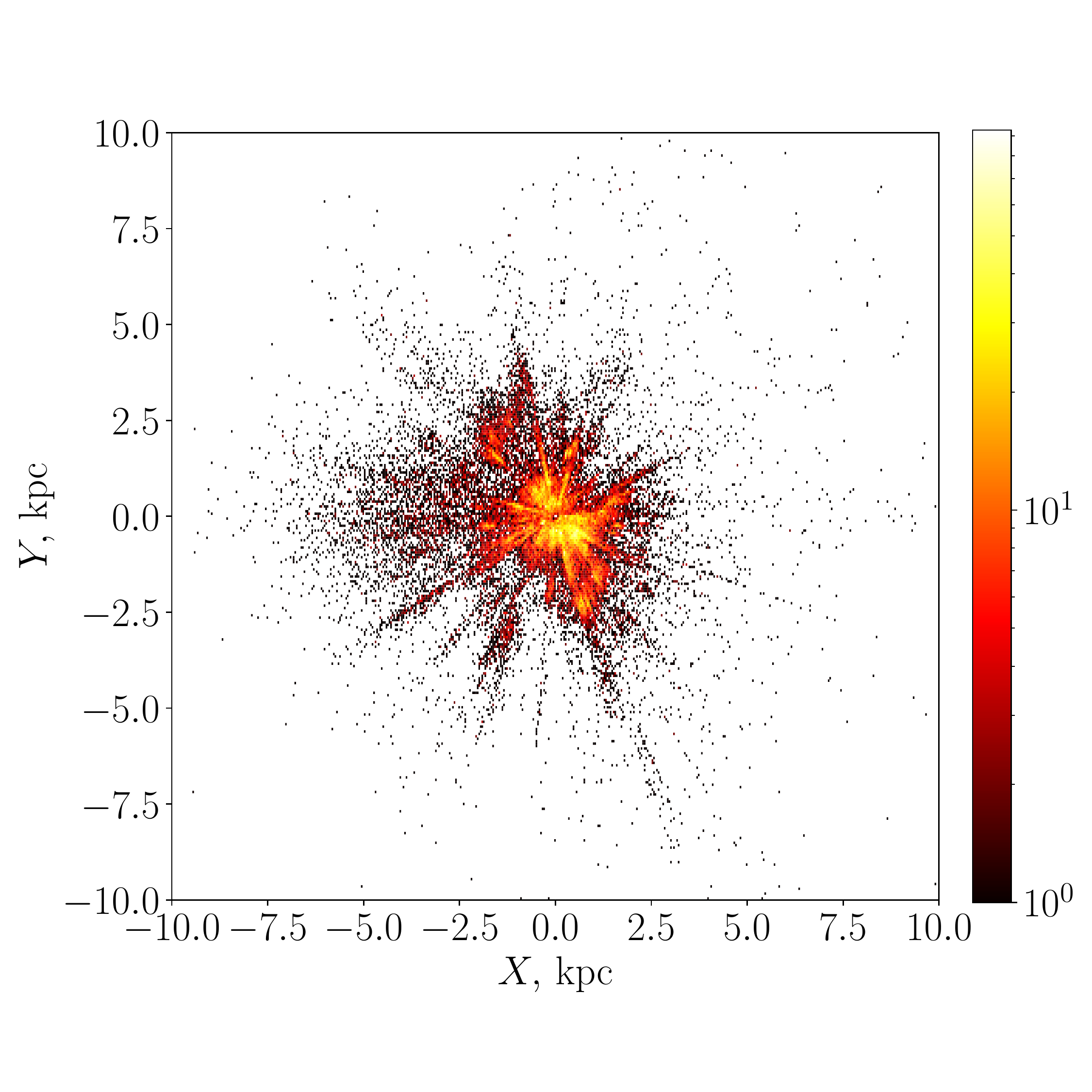}
	\caption{Distribution of our sample objects in the Galactic symmetry plane. The colour scale corresponds to the logarithm of their density. The $x$ axis is directed to the centre of the Galaxy, the $y$ axis is in the direction of Galactic rotation.}
	\label{fig:sample_galplane}
\end{figure}

The result was a sample of 47,735 objects, including 46,780 OB stars from various sources and 955 young OCs. Figures~\ref{fig:sample_sky} and \ref{fig:sample_galplane} show the distributions of our sample objects on the sky and in the Galactic symmetry plane, respectively.

\section{Algorithm}\label{algo}

At the initial stages of any data clustering task, one has to make choices as to the parameter space where the analysis will be performed and which clustering algorithm to adopt. We adopted the well-known HDBSCAN$^*$ algorithm, implemented in the \texttt{hdbscan} python library of the same name. This algorithm is well-suited for solving clustering problems in astronomical data, because it allows separating cluster members from `noise' (i.e., field stars), and it can also find clusters of any arbitrary shape. The latter aspect is relevant for astronomical applications, since errors in the distances, which are many orders of magnitude greater than errors in the (projected) positions, lead to significant elongations of stellar groups along the line of sight; not all clustering algorithms are equally good at finding highly elongated clusters. In addition, the HDBSCAN$^*$ algorithm is relatively easy to use, since it has only few input parameters, of which \textit{ min\_cluster\_size} is the main one. Its meaning is extremely intuitive, because it is simply the minimum cluster size one wants to derive from the final clustering step. An important feature of the HDBSCAN$^*$ algorithm which distinguishes it from the classical DBSCAN algorithm is a special procedure which allows searching for clusters throughout the condensed tree of the cluster hierarchy. In turn, this allows it to select clusters of variable density, whereas DBSCAN cuts this hierarchy at one specific density level (given as an additional parameter). Another useful HDBSCAN$^*$ option is the choice of method for selecting clusters from the cluster tree hierarchy. The user can choose between the `Excess of Mass' (EoM) and `leaf' methods. Our choice was the latter, because our goal is to search for objects of the same nature with similar properties, and the `leaf' method allows us to find more homogeneous clusters in the data, while maintaining the ability to cover a wide range of densities. The EoM, on the other hand, tends to select a small number of large clusters and a series of smaller groups in their agglomeration, which leads to the appearance in the final clustering of a number of physically improbable groups with sizes exceeding 1 kpc.

An optimal approach to search for stellar groups is to consider them in six-dimensional phase space, which includes three components of the spatial velocity and three spatial coordinates. However, the absence of metallic spectral lines and the small number and large widths of hydrogen and helium lines in the spectra of early-type stars make it difficult to obtain accurate estimates of their radial velocities. This is also evidenced by the fact that only 14 per cent of the sample stars have radial-velocity data according to the SIMBAD\footnote{\url{http://simbad.u-strasbg.fr/simbad/}} astronomical database, and only slightly more than half have radial velocities with a relative error of less than 50 per cent. Therefore, for our sample of young objects we excluded radial velocities from consideration at the clustering stage, thus limiting our phase space to five dimensions: two stellar proper-motion components, $\left (PM_{\alpha},\,PM_{\delta} \right)$, reflecting the components of the stars' transverse velocities and three coordinates $(X,\,Y,\,Z)$ in the heliocentric Cartesian frame associated with Galactic coordinates, calculated as follows:
\begin{equation}
\begin{split}
		X &= r_{\mathrm h} \cos l \cos b ;\\
	    Y &= r_{\mathrm h} \sin l \cos b ;\\ 
	    Z &= r_{\mathrm h} \sin b ,
\end{split}
\label{eq:cortesianxyz}
\end{equation}
where $\left( l, b \right)$ are Galactic longitude and latitude respectively and the stars' heliocentric distance $r_{\mathrm h}$ is geometric distance from \citet{BJ2021}. With this choice of coordinate system, $X$ is positive and oriented towards the Galactic centre, $Y$ is positive in the direction of Galactic rotation and $Z$ is positive towards the North Galactic Pole. We used the proper motions as proxies of the stellar velocity components, because transforming proper motions into linear velocities using $V_{\alpha, \delta} [\text{km s}^{-1}] = 4.741 r_{\mathrm h} [\text{kpc}] PM_{\alpha, \delta} [\text{mas yr}^{-1}]$ generates an additional distance-related uncertainty. 

Before applying the algorithm, we normalised all input parameters to unit variance so as to achieve greater uniformity of the data as a whole along the various coordinate axes and to facilitate the operation of the algorithm. Note that this procedure does not result in individual clusters with shapes close to spherical in the parameter space used, but thanks to the HDBSCAN$^*$ features, which can work with clusters of arbitrary shape, this is not fundamentally necessary.

\section{Results}\label{res}

\begin{figure}
	\includegraphics[width=\columnwidth]{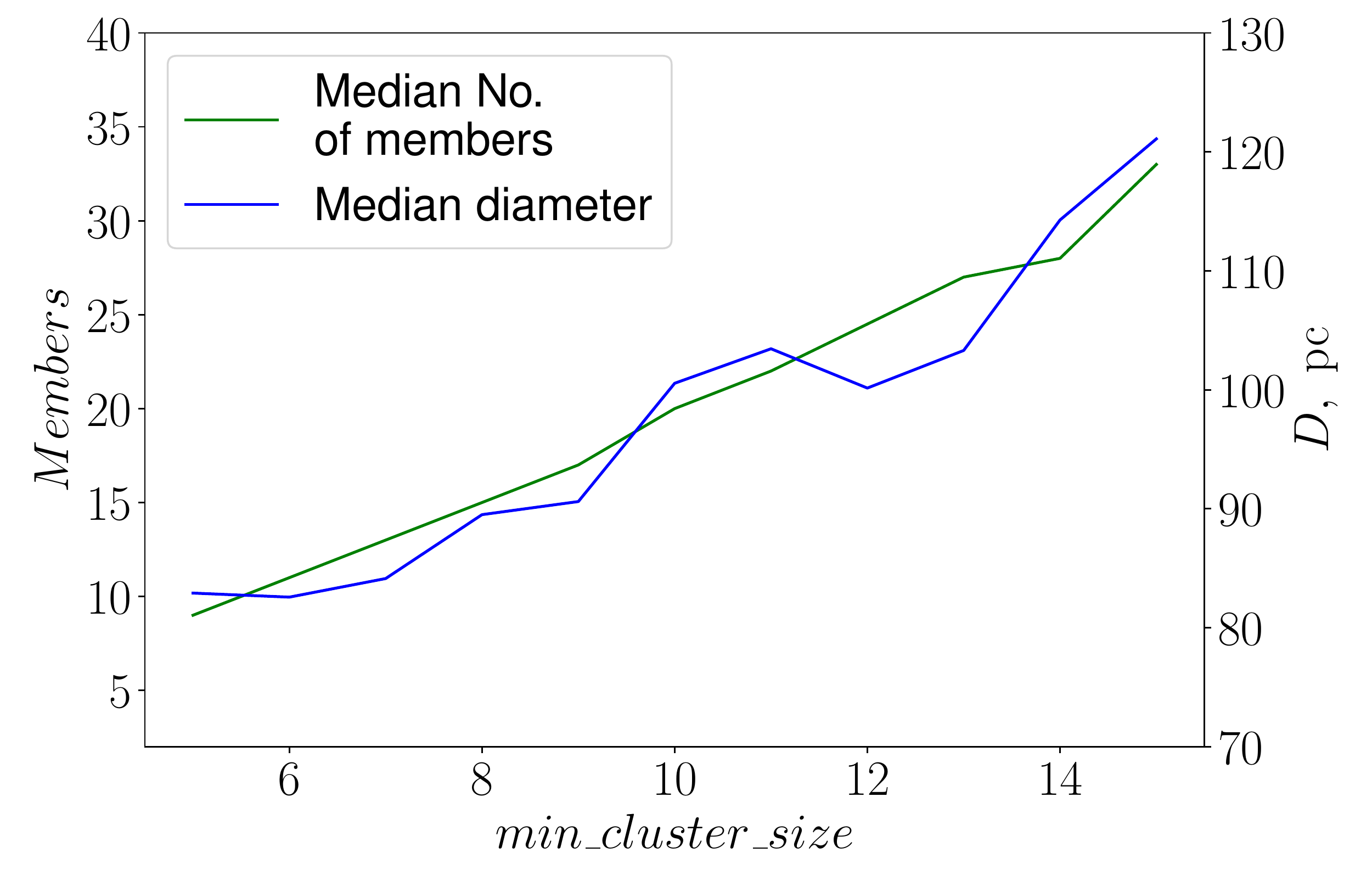}
	\caption{Dependence of the median number of members (green) and median diameter (blue) for the resulting clusters on \textit{min\_cluster\_size} in the range from 5 to 15.}
	\label{fig:min_clus_size}
\end{figure}

\begin{figure}
	\includegraphics[width=\columnwidth]{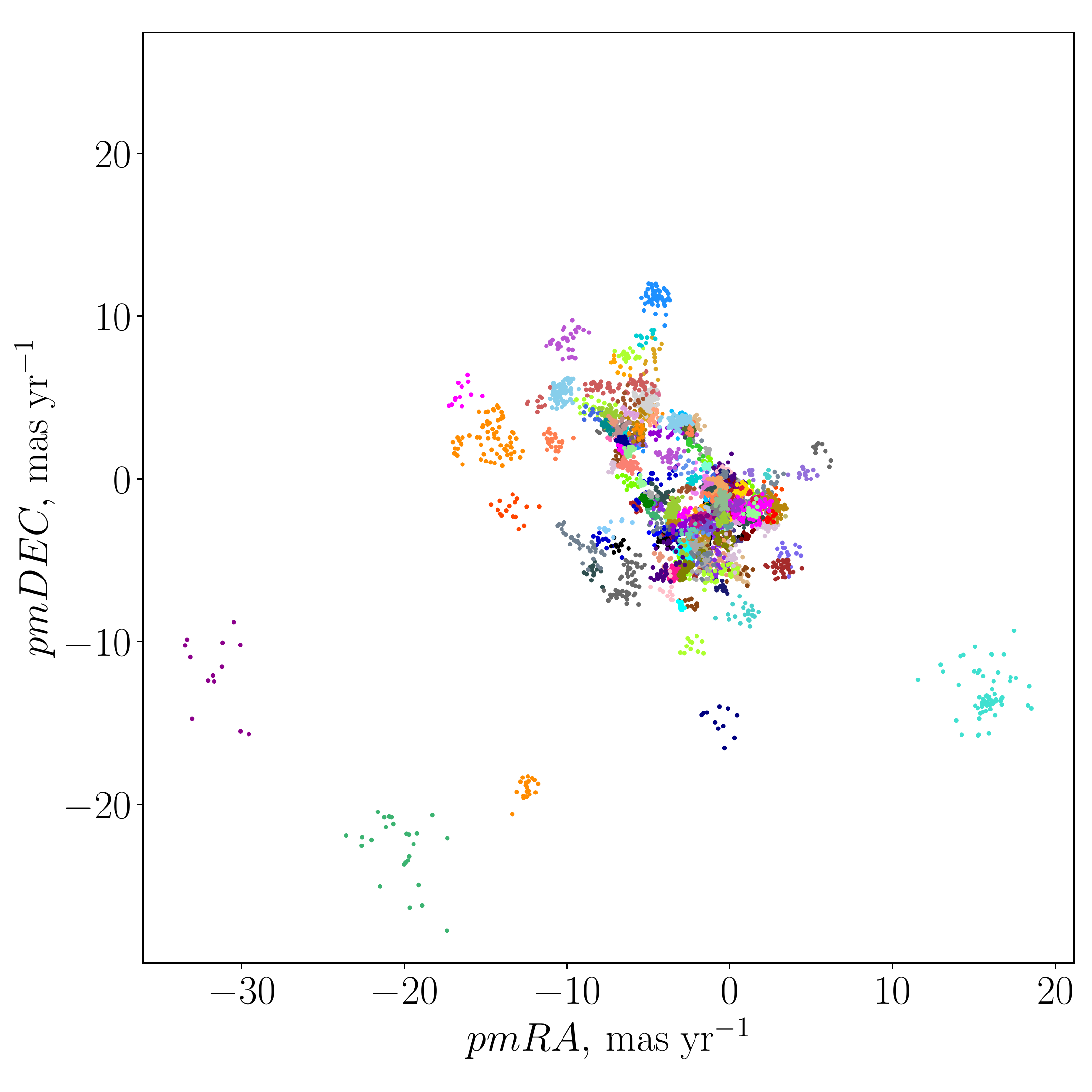}
	\caption{Clustering result for \textit{min\_cluster\_size} = 10 in proper-motion space. All 214 clusters are plotted using random colours.}
	\label{fig:clustersPM}
\end{figure}

\begin{figure}
	\includegraphics[width=\columnwidth]{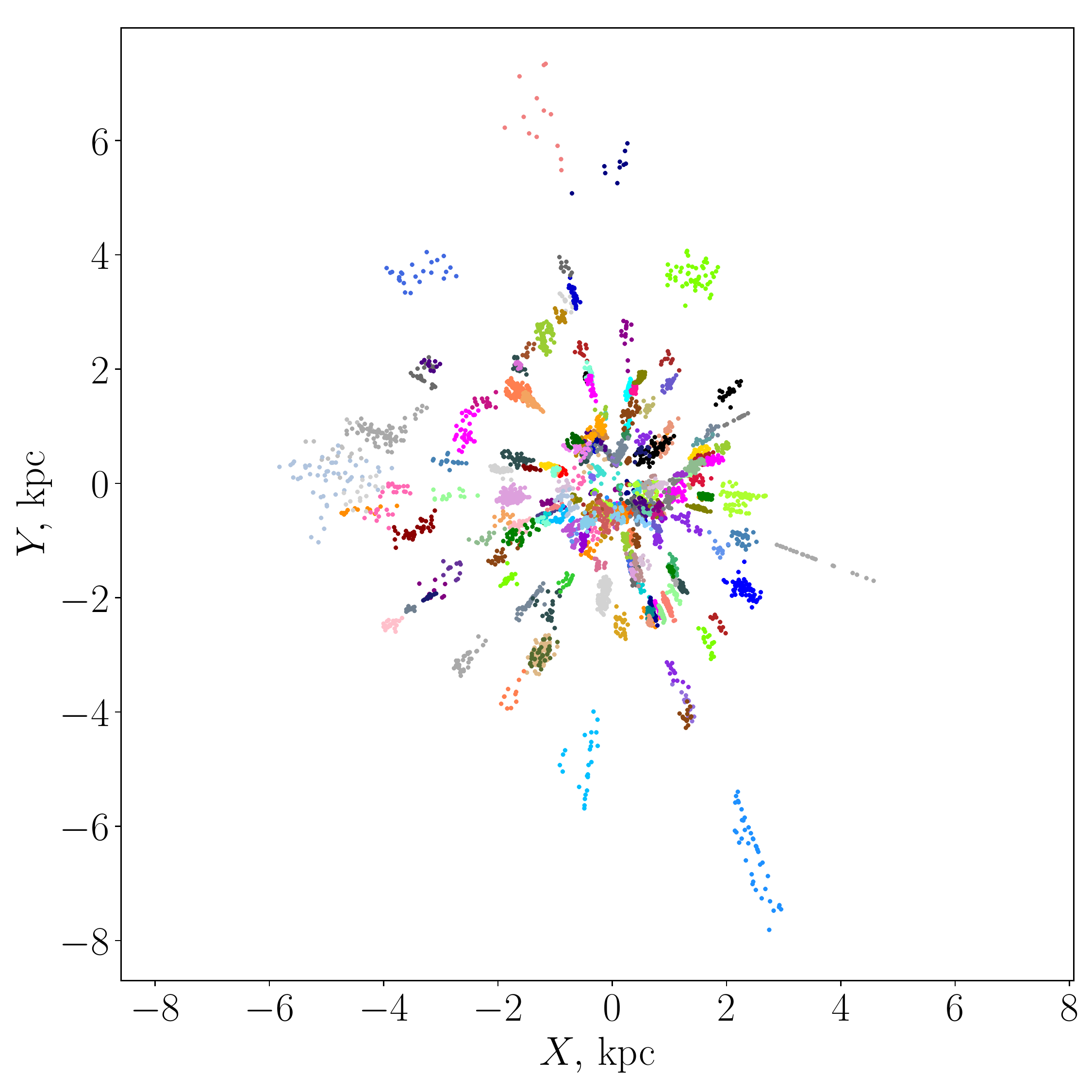}
	\caption{As Figure~\ref{fig:clustersPM}, but projected onto the Galactic plane. The colours correspond to the ones of the clusters in the Figure~\ref{fig:clustersPM}. The Galactic centre is on the right. The origin of coordinates corresponds to the position of the Sun.}
	\label{fig:clustersGalPlane}
\end{figure}

In order to carry out the clustering procedure, we used \textit{min\_cluster\_size} = 10. In the HDBSCAN$^*$ algorithm, \textit{min\_cluster\_size} is a relatively intuitive parameter to select; it represents the smallest size grouping one wishes to consider a cluster. In our choice we were guided, first, by considerations of the statistical significance of any clusters thus identified, given that the greater the number of objects in a cluster, the more likely it is not a result of random proximity of the parameters of several stars in a multidimensional parameter space. Second, we were driven by our desire to keep this value small enough to avoid losing many objects, because the number of clusters increases as the number of cluster members decreases, and a greater minimum cluster size leads to a significant reduction in the number of clusters identified. When estimating the kinematic ages of the clusters (see Section~\ref{ages}), we used the Monte-Carlo method, randomly choosing 90 per cent of cluster stars and removing 10 per cent. Hence, \textit{min\_cluster\_size} must be at least 10 (or greater), so that the formal number of stars removed at each iteration would not be less than one. We thus settled on \textit{min\_cluster\_size} = 10. Figure~\ref{fig:min_clus_size} demonstrates the dependence of the median number of members and the median diameter (the diameter for each cluster was estimated as the diameter of a circle containing 68 per cent of its members) for the resulting clusters on the \textit{min\_cluster\_size} value, in the range from 5 to 15. Both dependencies show an approximately linear increase with increasing \textit{min\_cluster\_size}. The characteristic size of clusters over the entire range considered remains within the values expected for such objects, mostly identified as OB associations.
Our clustering analysis of the sample using the HDBSCAN$^*$ algorithm with \textit{min\_cluster\_size} = 10 yielded a set of 214 groupings. The clusters found are shown in the proper-motion space and projected onto the Galactic plane in Figures~\ref{fig:clustersPM} and \ref{fig:clustersGalPlane}, respectively. 

The full versions of the tables containing the average cluster parameters, as well as lists of their members can be found as supplementary material. As an example, the Table~\ref{tab:example_avg_param} shows the parameters of clusters No. 34 -- 38 and the Table~\ref{tab:example_cluster64} lists the members of cluster No. 64.
\begin{table*}
	\centering
	\caption{The parameters of clusters No. 34 -- 38. No.: cluster number; $<l>$ and $<b>$: Galactic coordinates of the cluster center; $<PM_{\alpha}>$ and $<PM_{\delta}>$: average proper motions of cluster members; $<R_{\rm h}>$: average heliocentric distance to the cluster (based on geometric distances from \citet{BJ2021}); $\theta_{68}$ and $D_{68}$: the diameter of the region in the sky containing 68\% of the cluster members and the physical diameter of the cluster corresponding to the size of this region at the average heliocentric distance to the cluster, respectively; Memb: number of members in the cluster; $\sigma_V$ (corr): one-dimensional internal velocity dispersion that has undergone a correction procedure for effects not related to the direct motion of stars inside clusters (see Section~\ref{v_correstion}); $T_{\rm l}$: kinematic age of the cluster, estimated from the `proper motion--coordinate' diagram for Galactic longitude. Only positive age estimates, which do not exceed 200 Myr are indicated for clusters showing expansion at the one-sigma confidence level (see Section~\ref{ages}). Full version of this table with these and some additional parameters for all 214 clusters is available as supplementary material.}
	\label{tab:example_avg_param}
	\begin{tabular}{lcccccccccc} 
		\hline
		No. & $<l>$ & $<b>$ & $<PM_{\alpha}>$ & $<PM_{\delta}>$ & $<R_{\rm h}>$ & $\theta_{68}$ & $D_{68}$ & Memb & $\sigma_V$ (corr) & $T_{\rm l}$\\
		& deg & deg & mas\,yr$^{-1}$ & mas\,yr$^{-1}$ & pc & deg & pc & & km\,s$^{-1}$ & Myr\\
		\hline
            34 & 203.397 & 3.676 & -0.589 & -0.732 & 2418 & 2.23 & 94.21 & 14 & 4.21 & -- \\
            35 & 161.128 & -14.520 & 3.595 & -4.664 & 383 & 4.30 & 28.70 & 16 & 0.91 & 19.93 $\pm$ 0.87\\
            36 & 139.853 & -15.092 & 0.581 & -5.780 & 499 & 7.87 & 68.70 & 24 & 2.71 & 28.99 $\pm$ 0.12\\
            37 & 282.531 & -2.604 & -16.464 & 5.214 & 449 & 13.33 & 105.05 & 12 & 6.03 & -- \\
            38 & 76.131 & 1.516 & -2.370 & -7.699 & 1023 & 1.99 & 35.52 & 11 & 2.04 & 48.77 $\pm$ 1.79\\
		\hline
	\end{tabular}
\end{table*}
\begin{table*}
	\centering
	\caption{Members of cluster No. 64. ID: {\sl Gaia} EDR3 Source ID for stars and names for OCs; $\alpha,\,\delta$: J2016.0 equatorial coordinates; $\varpi$ and $\sigma \varpi$: {\sl Gaia} EDR3 parallax and uncertainty; $PM_{\alpha}$, $\sigma PM_{\alpha}$, $PM_{\delta}$ and $\sigma PM_{\delta}$: {\sl Gaia} EDR3 proper motions and uncertainties; $R_{\mathrm h}$: heliocentric geometric distance from \citep{BJ2021} with uncertainties; Class: object type (OB: O- and B-type stars, OPEN\_CLUS: OC). Full version of this table listing the members of all 214 clusters is available as supplementary material.}
	\label{tab:example_cluster64}
	\begin{tabular}{lcccccccccccccc} 
		\hline
		ID & label & $\alpha$ & $\delta$ & $\varpi$ & $\sigma \varpi$ & $PM_{\alpha}$ & $\sigma PM_{\alpha}$ & $PM_{\delta}$ & $\sigma PM_{\delta}$ & $R_{\mathrm h}$ & Class\\
		& & deg & deg & mas & mas & mas\,yr$^{-1}$ & mas\,yr$^{-1}$ & mas\,yr$^{-1}$ & mas\,yr$^{-1}$ & pc & \\
		\hline
          5616474861217929856 & 64 & 110.149 & $-$25.174 & 0.788 & 0.037 & $-$2.512 & 0.026 & 2.819 & 0.038 & 1267$_{-55}^{+73}$ & OB\\[3pt]
          5616650027161762688 & 64 & 110.688 & $-$25.171 & 0.840 & 0.044 & $-$2.158 & 0.028 & 2.794 & 0.045 & 1173$_{-47}^{+62}$ & OB\\[3pt]
          5617731431207979520 & 64 & 109.906 & $-$24.024 & 0.774 & 0.032 & $-$2.600 & 0.020 & 3.180 & 0.032 & 1275$_{-47}^{+48}$ & OB\\[3pt]
          5617293138386720768 & 64 & 109.402 & $-$24.816 & 0.742 & 0.036 & $-$3.024 & 0.026 & 2.989 & 0.035 & 1321$_{-59}^{+65}$ & OB\\[3pt]
          5616551552155481088 & 64 & 109.695 & $-$24.655 & 0.767 & 0.037 & $-$2.779 & 0.025 & 3.172 & 0.038 & 1291$_{-55}^{+69}$ & OB\\[3pt]
          5617719680177627136 & 64 & 110.025 & $-$24.086 & 0.811 & 0.025 & $-$2.820 & 0.016 & 3.004 & 0.027 & 1201$_{-27}^{+31}$ & OB\\[3pt]
          5616190225145335168 & 64 & 110.504 & $-$25.877 & 0.751 & 0.028 & $-$2.559 & 0.021 & 2.720 & 0.029 & 1313$_{-48}^{+37}$ & OB\\[3pt]
          5617731431207980288 & 64 & 109.903 & $-$24.023 & 0.836 & 0.028 & $-$2.633 & 0.016 & 3.105 & 0.028 & 1174$_{-41}^{+38}$ & OB\\[3pt]
          5616531142471195264 & 64 & 109.702 & $-$24.949 & 0.770 & 0.044 & $-$2.681 & 0.031 & 2.902 & 0.049 & 1272$_{-80}^{+65}$ & OB\\[3pt]
          5616525232596250624 & 64 & 109.675 & $-$24.970 & 0.821 & 0.030 & $-$2.956 & 0.022 & 2.958 & 0.032 & 1197$_{-35}^{+44}$ & OB\\[3pt]
          5616537292864419328 & 64 & 109.654 & $-$24.935 & 0.776 & 0.030 & $-$2.015 & 0.021 & 2.700 & 0.029 & 1255$_{-38}^{+58}$ & OB\\[3pt]
          NGC\_2362 & 64 & 109.671 & $-$24.949 & 0.743 & 0.077 & $-$2.779 & 0.176 & 2.950 & 0.231 & 1233$_{-57}^{+57}$ & OPEN\_CLUS\\[3pt]
		\hline
	\end{tabular}
\end{table*}

Table~\ref{tab:clustering_stats} lists the statistical parameters obtained from the clustering analysis.
\begin{table}
	\centering
	\caption{Statistical parameters obtained for the resulting clustering. The diameters of clusters were estimated as the diameter of a circle projected onto the celestial sphere, containing 68 per cent of the members of a given cluster and located at a distance equal to the average heliocentric distance to the cluster. Angular sizes represent these diameters in degrees. `Raw' values of dispersion were calculated directly from the parameters of the cluster members. `Corrected' values were calculated after correcting the one-dimensional velocity dispersions for effects not directly related to the internal motions of the cluster members (see Section~\ref{v_correstion}). The width of the distribution is represented by the Median Absolute Deviation (MAD), which is much less sensitive to outliers in the tails of the distribution than the standard deviation.}
	\label{tab:clustering_stats}
	\begin{tabular}{lcc} 
		\hline
		Parameter & Median & MAD\\
		\hline
		Angular size, $\degr$ & 4.17 & 2.30\\
		Diameter, pc & 98.7 & 42.4\\
		Number of members & 20 & 8\\
		$\sigma_V$ (raw), km\,s$^{-1}$ & 2.38 & 0.91\\
		$\sigma_V$ (corrected), km\,s$^{-1}$ & 2.36 & 0.94\\
		\hline
	\end{tabular}
\end{table}

\subsection{Velocity dispersion correction}\label{v_correstion}
We corrected the one-dimensional velocity dispersions of the resulting clusters for two important effects:
\begin{enumerate}
	\item Proper-motion errors, which increase the observed dispersion of the transverse velocities;
	\item Nonzero radial velocities, which may result in the appearance of additional stellar velocity components in the plane of the sky and which are not related to their motion in that plane: an association moving away from (approaching) us seems to be contracting (expanding), since the stellar velocity vectors seem to be directed towards (away from) the apex point of the association's motion.
\end{enumerate}

To take into account the former effect, the component associated with the average uncertainty of the proper motions along the corresponding coordinate was subtracted from the association's velocity dispersion. We adopted the proper-motion uncertainties from {\sl Gaia} EDR3, $\langle \text{err}(PM_{l}) \rangle_{\mathrm{sample}} = 0.022 \text{ mas yr}^{-1}$ and $\langle \text{err}(PM_{b}) \rangle_{\mathrm{sample}} = 0.023 \text{ mas yr}^{-1}$ (representing averages for the entire sample).

To correct the velocity dispersion for the motion of an association as a whole, along the line of sight, we used the recipe of \citet{Vaher2020} to calculate the transverse components of the velocities, $\tilde{V}_{l},\,\tilde{V}_{b}$, of cluster members in the cluster-centred reference frame. This procedure is good in that it allows one to calculate the components of the stellar velocities even in the presence of a very rough estimate of their radial velocities. As alluded to already, there is very little accurate information about the radial velocities of stars of early spectral types owing to their spectral features. Therefore, as a rough value of the radial velocities of individual stars, we attributed the average radial velocity of the cluster itself to all cluster members. This latter parameter was calculated in two different ways: if we could estimate the average observed radial velocity of a cluster based on data from the SIMBAD catalogue, we used it; if use of such an estimate was not possible owing to a small number of stars with known radial velocities or significant scatter of these radial velocities, we used model radial velocities obtained under the assumption of circular cluster orbits around the Galactic centre, because most OB stars have orbits close to circular \citep{BoBa2019}. The observed mean radial velocity was used if it was calculated from at least five stars and if its uncertainty did not exceed 5 km~s$^{-1}$. 

The model radial velocity used otherwise was calculated as follows. In the approximation of circular cluster orbits, the clusters' heliocentric radial velocities can be calculated as one of the components of their total model velocity,
\begin{equation}
V_{\mathrm{model, assoc}} = \left(
\begin{array}{c}
V_{r}\\
V_{l}\\
V_{b}
\end{array}
\right) = G \times 
\left(
\begin{array}{c}
U_{0}\\
V_{0}\\
W_{0}
\end{array}
\right) + V_{\mathrm{diff}},
\label{eq:modelV}
\end{equation}
where 
\begin{equation}
G = \left(
\begin{array}{ccc}
\cos b_{0} \cos l_{0} & \cos b_{0} \sin l_{0} & \sin b_{0}\\
- \sin l_{0} & \cos l_{0} & 0\\
- \sin b_{0} \cos l_{0} & - \sin b_{0} \sin l_{0} & \cos b_{0}
\end{array}
\right)
\label{eq:Gmatrix}
\end{equation}
and
\begin{equation}
V_{\mathrm{diff}} = \left(
\begin{array}{c}
R_0 \left( \omega - \omega_0 \right) \sin l_0 \cos b_0 \\
\left( R_0 \cos l_0 - r_{\mathrm h} \cos b_0 \right) \left( \omega - \omega_0 \right) - r_{\mathrm h} \omega_0 \cos b_0 \\
- R_0 \left( \omega - \omega_0 \right) \sin l_0 \sin b_0.
\end{array}
\right)
\label{eq:Vdiff}
\end{equation}

Here, $R_0 = 8.21 \text{ kpc}$ (\citet{RastUt17}, based on kinematics of Galactic masers; \citet{deGrijsBono2016} based on statistical analysis) is the distance from the Sun to the centre of the Galaxy. $U_0 = -11.06 \text{ km s}^{-1}, V_0 = -18.26 \text{ km s}^{-1}, \text{ and } W_0 = -8.76 \text{ km s}^{-1}$ \citep{RastUt17} are the velocity components of the local sample relative to the Sun. We calculated the angular velocity of the Galactic disc at a given Galactocentric radius, $\omega (r_{\mathrm g}) = \omega (\sqrt{R_0^2 + r_{\mathrm h}^2 \cos^2 b_0 - 2 R_0 r_{\mathrm h} \cos b_0 \cos l_0})$ by expanding it into a Taylor series up to and including the fourth order,
\begin{equation}
\begin{split}
\omega (r_{\mathrm g}) &= \omega_0 + \omega_0' (r_{\mathrm g} - R_0) + \frac{1}{2!} \omega_0'' (r_{\mathrm g} - R_0)^2 +\\
&+ \frac{1}{3!} \omega_0''' (r_{\mathrm g} - R_0)^3 + \frac{1}{4!} \omega_0'''' (r_{\mathrm g} - R_0)^4,
\end{split}
\label{eq:omegaTailor}
\end{equation} 
where $r_{\mathrm g}$ is the association's Galactocentric radius. We adopted the model parameters $R_0, U_0, V_0, W_0$, as well as the expansion coefficients of the angular velocity, $\omega_0 = 28.94 \text{ km} \text{ s}^{-1} \text{ kpc}^{-1}$, $\omega_0' = -3.91 \text{ km} \text{ s}^{-1} \text{ kpc}^{-2}$, $\omega_0'' = 0.86 \text{ km} \text{ s}^{-1} \text{ kpc}^{-3}$, $\omega_0''' = 0.01 \text{ km} \text{ s}^{-1} \text{ kpc}^{-4}$, $\omega_0'''' = -0.08 \text{ km} \text{ s}^{-1} \text{ kpc}^{-5}$ from \citet[][model A1]{RastUt17}, which those authors estimated based on kinematic analysis of the Galactic maser sample.

In view of the foregoing, the expression for calculating the velocity components in the cluster-centred reference frame has the following form:
\begin{equation}
\left(
\begin{array}{c}
\tilde{V}_{l}\\
\tilde{V}_{b}\\
\tilde{V}_{r}
\end{array}
\right) = \mathbf{A} \times 
\left(
\begin{array}{c}
V_{l}\\
V_{b}\\
V_{r,0}
\end{array}
\right) - 
\left(
\begin{array}{c}
V_{l,0}\\
V_{b,0}\\
V_{r,0}
\end{array}
\right)
\label{eq:centredV}
\end{equation}
where
\begin{equation}
\mathbf{A} = 
\left(
\begin{array}{ccc}
a_{1,1} & a_{1,2} & a_{1,3}\\
a_{2,1} & a_{2,2} & a_{2,3}\\
a_{3,1} & a_{3,2} & a_{3,3}
\end{array}
\right)
\label{eq:matrixA}
\end{equation}
\begin{equation}
\begin{array}{l}
a_{1,1} = \cos(l - l_0)\\
a_{1,2} = -\sin b \sin(l - l_0)\\
a_{1,3} = \cos b \sin(l - l_0)\\
a_{2,1} = \sin b_0 \sin(l - l_0)\\
a_{2,2} = \cos b_0 \cos b + \sin b_0 \sin b \cos(l - l_0)\\
a_{2,3} = \cos b_0 \sin b - \sin b_0 \cos b \cos(l - l_0)\\
a_{3,1} = -\cos b_0 \sin(l - l_0)\\
a_{3,2} = \sin b_0 \cos b - \cos b_0 \sin b \cos(l - l_0)\\
a_{3,3} = \sin b_0 \sin b + \cos b_0 \cos b \cos(l - l_0)
\end{array}
\label{eq:matrixAcomp}
\end{equation}
Here, index 0 denotes a value related to the cluster centre.

The final expression for the one-dimensional velocity dispersion in the plane of the sky, corrected for errors in proper motions and the overall movement of the association along the line of sight, has the following form (quantities with a tilde denote corrected values):
\begin{equation}
\begin{split}
&\sigma_{\mathrm{total}} \left( V_{l} \right) = \sqrt{ \sigma^2  \left( \tilde{V}_{l} \right) - \langle err_{\mathrm{PM_{l}}} \rangle_{\mathrm{assoc}}^2 \left( 4.741 r_{\mathrm h} \right)^2}\\
&\sigma_{\mathrm{total}} \left( V_{b} \right) = \sqrt{ \sigma^2  \left( \tilde{V}_{b} \right) - \langle err_{\mathrm{PM_{b}}} \rangle_{\mathrm{assoc}}^2 \left( 4.741 r_{\mathrm h} \right)^2}\\
&\sigma_{\mathrm{total}} = \sqrt{0.5 \sigma_{\mathrm{total}}^2 \left( V_{l} \right) + 0.5 \sigma_{\mathrm{total}}^2 \left( V_{b} \right)},
\end{split}
\label{eq:finalSigmaV}
\end{equation}
where $(\sigma (\tilde{V}_{l}), \sigma (\tilde{V}_{b}))$ are the velocity dispersions for $(\tilde{V}_{l}, \tilde{V}_{b})$ inside the association, and $(\langle err_{\mathrm{PM_{l}}} \rangle_{\mathrm{assoc}}, \langle err_{\mathrm{PM_{b}}} \rangle_{\mathrm{assoc}})$ are the average errors of the proper motions of the association along the $(l, b)$ coordinates.

It is known that in the solar neighbourhood there are significant velocity-field irregularities, whose presence can lead to perceptible errors in the estimation of radial velocities in the framework of the circular orbits model. In order to assess whether in our case the use of radial velocities directly from the rotation curve is justified, we performed a simulation using the Monte Carlo method. \citet{MelDam2009} give the standard deviation of the velocities of OB associations from the rotation curve, $\sigma_{\mathrm{assoc}} = 7.2$ km s$^{-1}$. For each cluster, we computed a one-dimensional velocity dispersion, uncorrected for motion along the line of sight $\sigma_{0}$ (i.e. calculated for a value $V_{\rm r} = 0$ km s$^{-1}$), a similar value with a line-of-sight velocity from the rotation curve $\sigma_{\rm{RC}}$ and a set of 100 velocity dispersion estimates $\sigma_{\rm{rand}}$ computed from 100 Gaussian-distributed random radial velocities around the velocity from the rotation curve with standard deviation of 7.2 km s$^{-1}$. For each cluster, let $\Delta_1$ be the difference between the uncorrected velocity dispersion and one corrected in accordance with the rotation curve, i.e. $\Delta_1 = | \sigma_{\rm{RC}} - \sigma_{0} |$. This value characterises the error we introduce without taking into account the effect of radial velocities at all. Let the value $\Delta_2$ for each cluster be the standard deviation of the sample of 100 random velocity dispersion estimates obtained, i.e. $\Delta_1 = STD (\sigma_{\rm{rand}})$. This value, in turn, characterises the error in estimating the internal velocity dispersion, which is generated by possible deviations of the kinematics of the OB associations from the rotation curve. On average across the entire sample of 214 final clusters, $\langle \Delta_1 \rangle = 0.164$ is significantly larger than $\langle \Delta_2 \rangle = 0.111$. Thus, the errors resulting from the deviations in association with the kinematics from the rotation curve, which arise when a simple circular orbit model is applied, are smaller than the error that would be introduced if this effect were ignored.

\subsection{Kinematic ages}\label{ages}
\begin{figure}
	\includegraphics[width=\columnwidth]{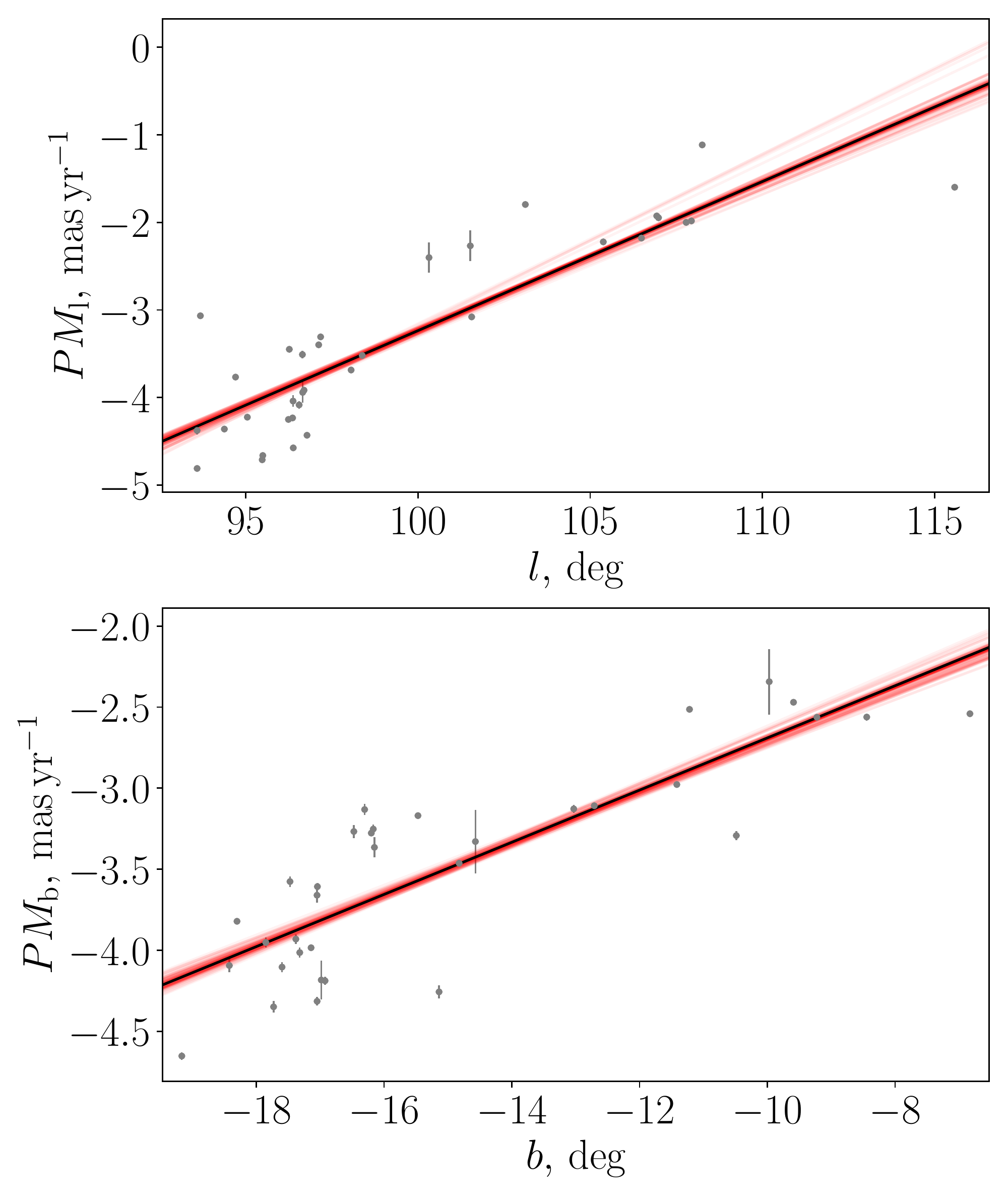}
	\caption{(Top) $(PM_{\mathrm l}, l)$ diagram for cluster No. 44. (Bottom) $(PM_{\mathrm b}, b)$ diagram for the same cluster. Straight black lines represent linear fits for median kinematic ages of $21.15 \pm 0.27$ Myr and $22.37 \pm 0.42$ Myr for the top and bottom panels, respectively. The red solid lines show the spread of the results of the least-squares method from all 100 iterations.}
	\label{fig:clus44_diagrams}
\end{figure}

To determine the kinematic ages of the resulting clusters, we examined the `proper motion--coordinate' diagrams in Galactic longitude and latitude. The key expansion marker is the presence of a correlation between the coordinate and the corresponding proper motion (an anti-correlation indicates contraction). We used the Monte Carlo method, as follows: 90 per cent of cluster stars were randomly selected each time, and a kinematic age estimate was determined for the  subsample thus obtained. This procedure was repeated 100 times. At each iteration, the least-squares method was used to estimate the age: `proper motion--coordinate' diagrams were approximated by a linear function of the form $PM_{j} = k_{j}(j - j_0) + c_{j}$, where $j = (l, b)$. Inverse velocity gradients provide an estimate of the kinematic age of a cluster, i.e., the period during which the expansion occurred. Note that the ages thus obtained are upper limits, since cluster expansion does not begin from a point but from some finite size and, therefore, it will last less long. For small clusters, discarding 10 per cent of their stars can significantly affect the results for the individual iterations, so strong deviations in the final age estimates cannot be ruled out. Therefore, we derived the final age estimate as the median of 100 iterations and used the median absolute deviation (MAD) as an estimate of its uncertainty, as it is less affected by outliers than the mean and standard deviation. Figure~\ref{fig:clus44_diagrams} shows an example of $(PM_{l}, l)$ and $(PM_{b}, b)$ diagrams with the corresponding linear fits for cluster No. 44, which can probably be identified with the Lacerta (LAC) OB1 association.

To control the quality of the correlations between proper motions and the corresponding coordinates, we used the Student's $t$-test. The observed values of the $t$ parameter for assessing the significance of a correlation can be found as follows:
\begin{equation}
t_{\mathrm{obs}} = \sqrt{\frac{r^2_{\mathrm{y,x}}}{1 - r^2_{\mathrm{y,x}}}(n - 2)},
\label{eq:t_obs}
\end{equation}
where $n$ is the number of points and $r_\mathrm{y,x}$ is the correlation coefficient, which by definition is equal to
\begin{equation}
r_\mathrm{y,x} = \frac{\sum\limits_{i=1}^n (x_{\mathrm i} - \langle x \rangle) \cdot (y_{\mathrm i} - \langle y \rangle)}{\sqrt{\sum\limits_{i=1}^n (x_{\mathrm i} - \langle x \rangle)^2 \cdot \sum\limits_{i=1}^n (y_{\mathrm i} - \langle y \rangle)^2}}.
\label{eq:corr_coef}
\end{equation}

Values of $t_\mathrm{obs}$ are distributed according to the Student's $t$ distribution with $(n - 2)$ degrees of freedom. We calculated the theoretical values $t_{\mathrm{theor}}$ of the parameter $t$ corresponding to confidence levels of 1$\sigma$ (68.3 per cent), 2$\sigma$ (95.4 per cent) and 3$\sigma$ (99.7 per cent) of the Gaussian distributions using the statistical software package \texttt{SciPy},\footnote{\url{https://scipy.org/}} in Python. The realisation of the condition $t_\mathrm{obs} \geq t_{\mathrm{theor}}$ means that the `proper motion--coordinate' diagram for this cluster demonstrates the presence of a correlation (and, hence, a sign of overall expansion) at the level of confidence for which the value $t_{\mathrm{theor}}$ was determined.

\begin{figure}
	\includegraphics[width=\columnwidth]{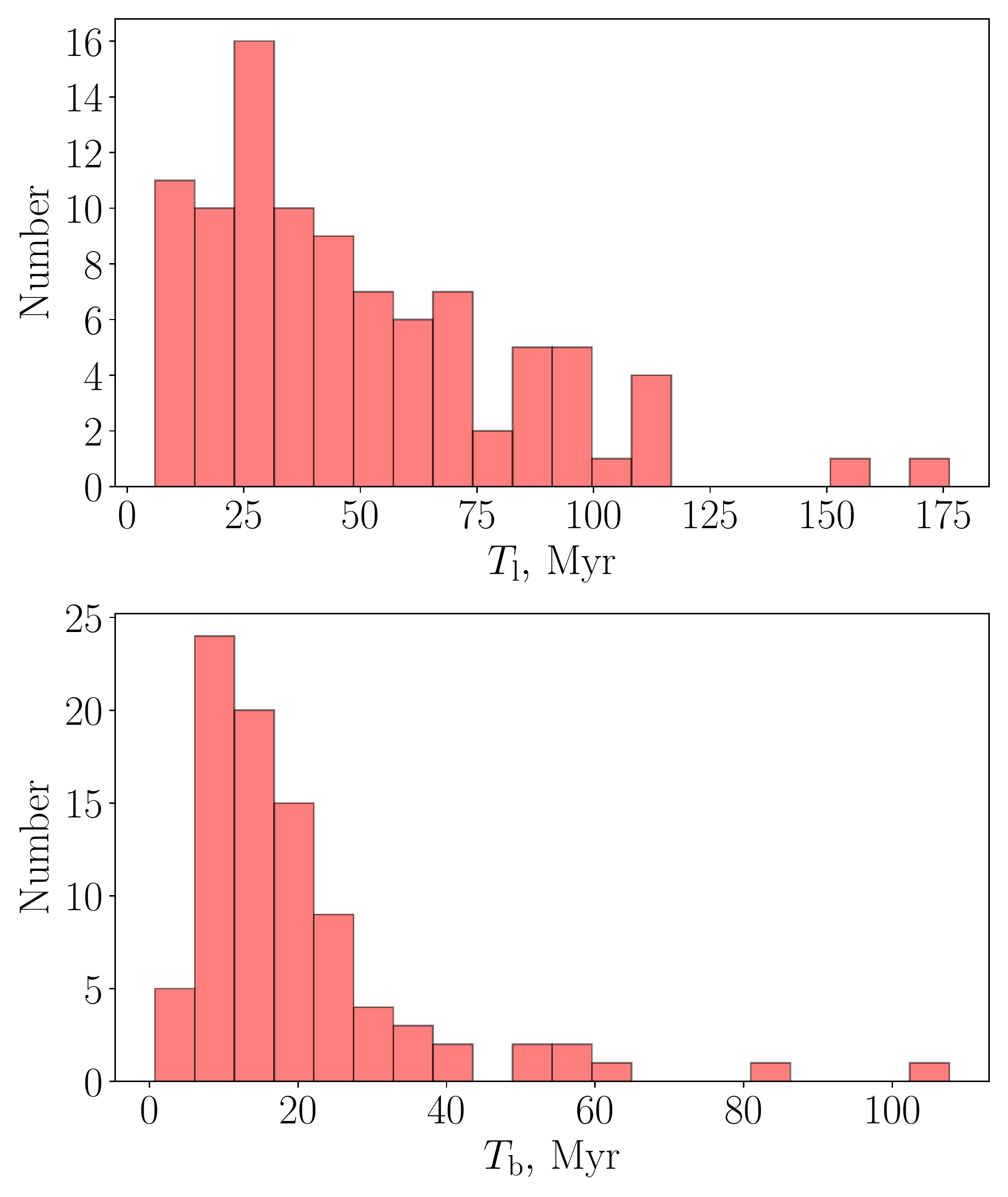}
	\caption{Distribution of ages obtained from `proper motion--coordinate' diagrams for Galactic longitude (top) and latitude (bottom) for clusters that satisfied the Student's $t$ test at the 1$\sigma$ confidence level.}
	\label{fig:age_hists}
\end{figure}

\begin{figure}
	\includegraphics[width=\columnwidth]{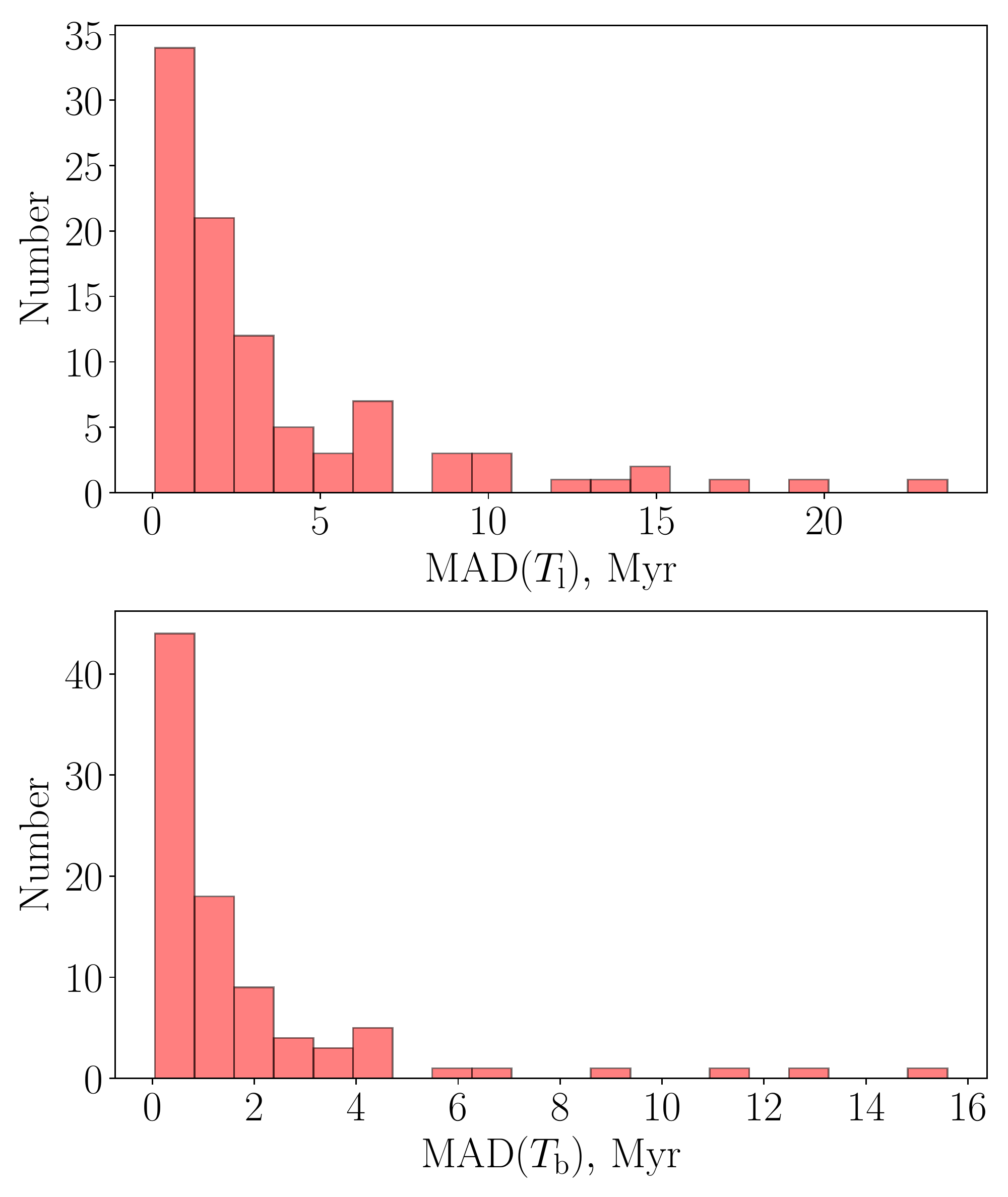}
	\caption{As Figure~\ref{fig:age_hists} but for the age uncertainties, defined as median absolute deviation (MAD)}
	\label{fig:e_age_hists}
\end{figure}

Clusters with negative age estimates were not taken into account for reasons of non-physicality. We also excluded clusters whose ages exceeded 200 Myr, since stellar associations can hardly be preserved for such a long time; the characteristic age of known associations does not exceed several tens of Myr. Thus, according to these criteria, 95 clusters show a correlation at the 1$\sigma$ level of confidence in longitude and 89 in latitude (47 in both); 38 clusters show a correlation at the 3$\sigma$ level of confidence in longitude and 25 in latitude (11 in both). Figures~\ref{fig:age_hists} and \ref{fig:e_age_hists} show the distributions of the ages of those clusters which satisfied the Student's $t$-test at the 1$\sigma$ confidence level and the distributions of their uncertainties, respectively.

The lack of older $T_{\mathrm b}$ ages among these clusters can be explained as follows. Movement of objects perpendicular to the Galactic disc is strongly affected by vertical oscillations. The period of vertical oscillations in the solar vicinity is about 87 Myr \citep{BinTre2nd}, based on the volume density of the Galactic disc in the solar neighbourhood, $\rho = 0.10\,M_{\sun} \mathrm{~pc}^{-3}$ \citep{Flynn2006}. On time-scales comparable with a quarter of this period, the assumption of linear motion of the stars loses validity and so the more complex nature of motions in the Galactic disc should be taken into account to determine cluster ages. In this regard, the quality of the diagrams deteriorates and the linear relationship of proper motions and coordinates ceases to manifest itself. The quality of the linear correlation between the latter parameters is declining, and we are increasingly less likely to encounter clusters that demonstrate the presence of such a correlation at a high level of confidence. This clearly demonstrates that the number of clusters with $T_{\mathrm b} \geq 20$--25 Myr (which approximately corresponds to a quarter of the period of vertical oscillations) begins to drop steeply. With an increase in the level of confidence, this effect is getting stronger, and among those clusters that satisfy the $t$ criterion at the 3$\sigma$ level for latitude, only one has a $T_{\mathrm b}$ age of more than 25 Myr. Since most $T_{\mathrm b}$ age estimates are less reliable than $T_{\mathrm l}$, we will henceforth focus on the $T_{\mathrm l}$ ages.

\begin{figure}
	\includegraphics[width=\columnwidth]{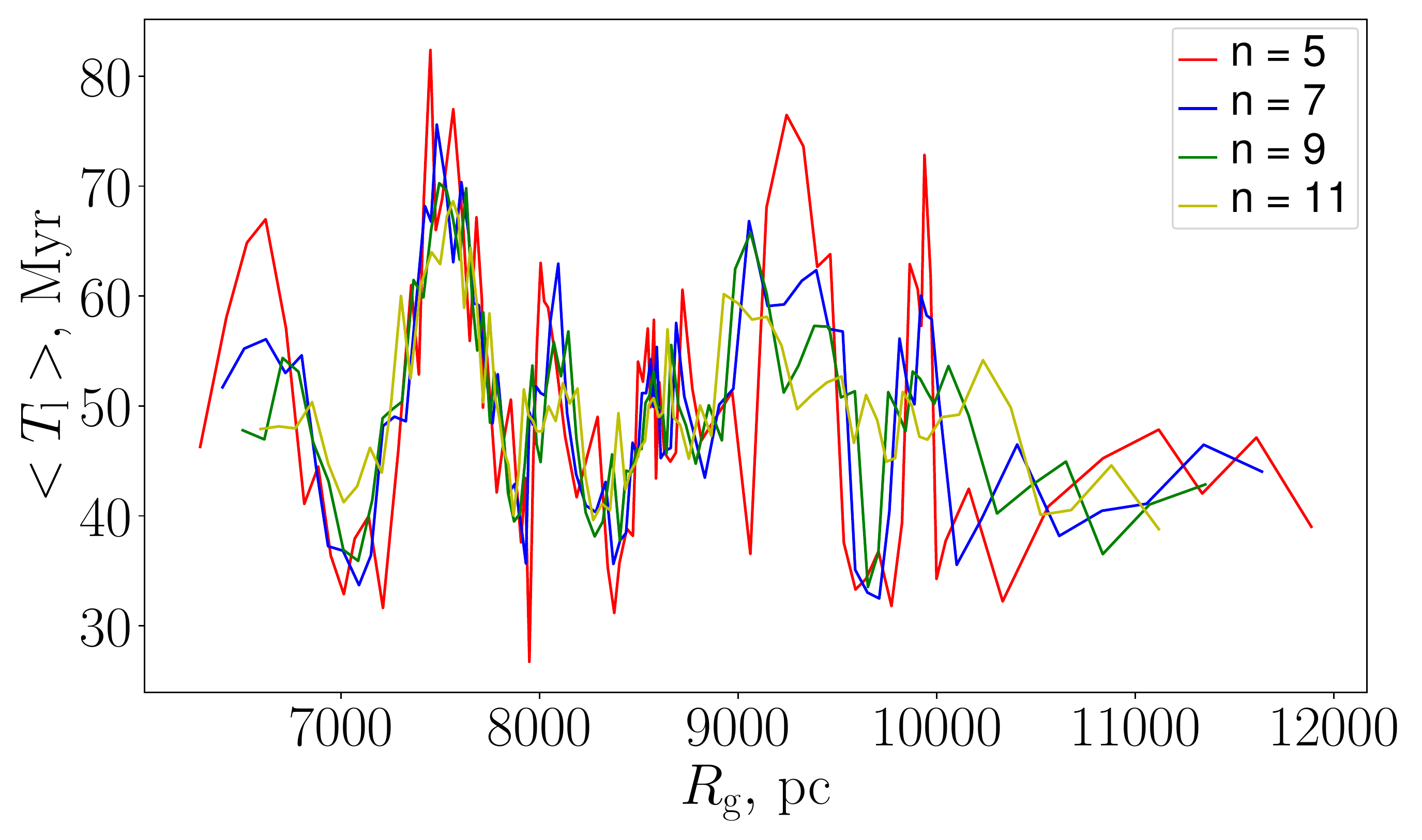}
	\caption{Smoothing of the $(T_{\mathrm l}, R_{\mathrm g})$ dependence using a moving average over $n =$ 5, 7, 9 and 11 data points.}
	\label{fig:rg_age}
\end{figure}

\begin{figure}
	\includegraphics[width=\columnwidth]{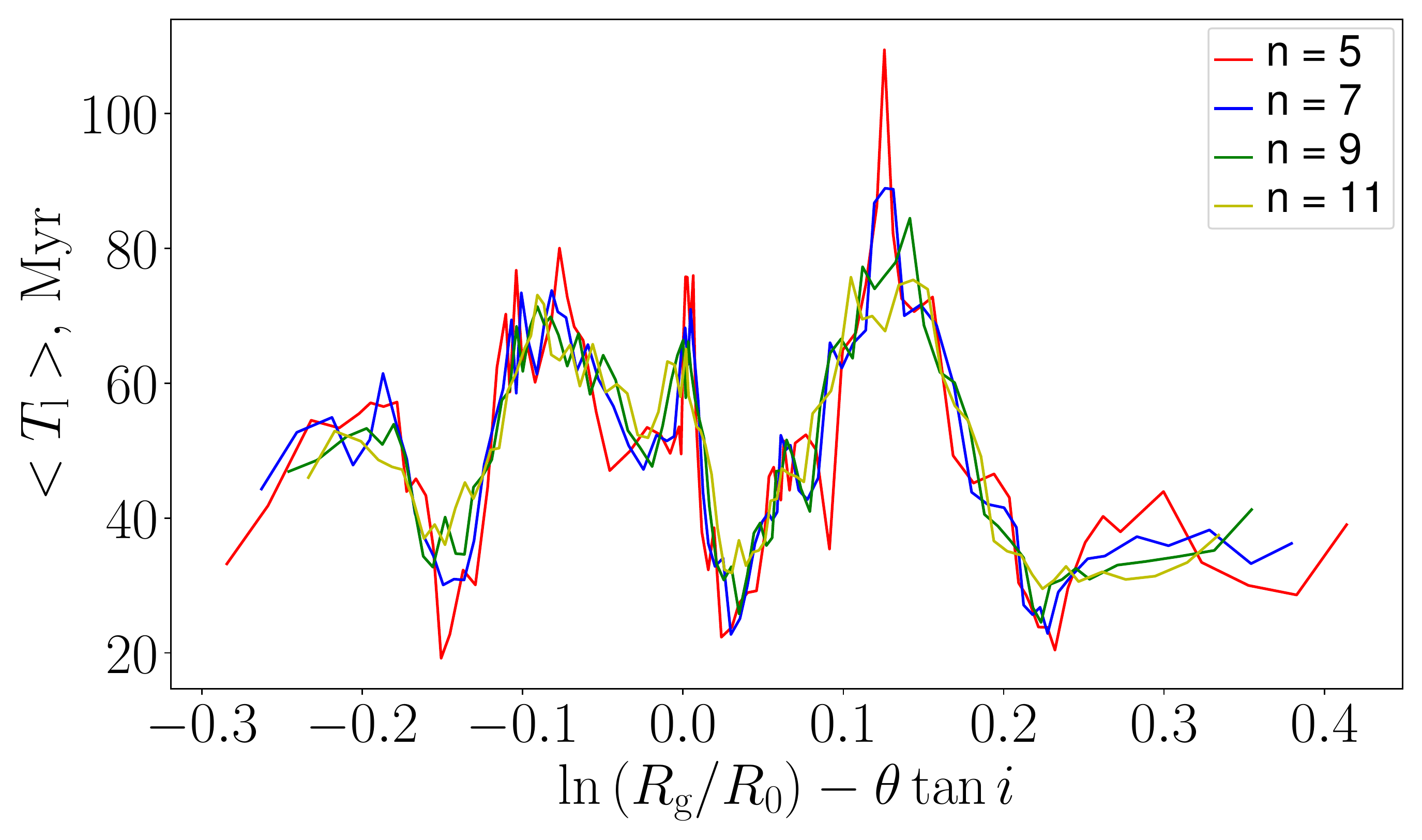}
	\caption{As Figure~\ref{fig:rg_age}, but for the parameter $\xi = \ln{(R_{\mathrm g} / R_0)} - \theta \tan(i)$ instead of the Galactocentric distance $R_{\mathrm g}$}
	\label{fig:rspir_age}
\end{figure}

We also examined the spatial distribution of clusters of various ages. We adopted the kinematic age estimates derived from Galactic longitude expansion as most reliable, which thus limited us to 95 clusters which show expansion at the 1$\sigma$ confidence level. In order to highlight possible trends in the dependence of age on the distance to the Galactic centre, we applied a series of smoothing kernels to the $(T_{\mathrm l}, R_{\mathrm g})$ dependence, using a moving averages over $n =$ 5, 7, 9 and 11 data points, and displayed the average dependencies thus obtained in Figure~\ref{fig:rg_age}. Dependencies obtained on all smoothing scales generally follow each other and do not demonstrate a clear large-scale trend between age and Galactocentric distance, but they exhibit a series of peaks and depressions.
To establish the nature of this `wavy' behaviour, we carried out a similar procedure, using instead of the Galactocentric distance the $\xi = \ln{(R_{\mathrm g} / R_0)} - \theta \tan(i)$ parameter, which allows us to examine the positions of objects relative to the spiral pattern described by logarithmic spirals. Here $R_0 = 8210$ pc is the solar Galactocentric distance, $\theta$ the Galactocentric azimuth and $i = -10.\degr4$ the pitch angle of the spiral pattern \citep[see ][]{RastUt17}. The result is shown in Figure~\ref{fig:rspir_age}.
Figure~\ref{fig:rspir_age} shows a clear correlation between the position of the peaks and the spiral arms of the Galaxy. The first peak ($-0.3 \leq \xi \leq -0.15$) captures a part of the Inner arm, the second peak ($-0.15 \leq \xi \leq 0.02$) is in good agreement with the position of the Carina--Sagittarius (Car--Sag) arm and the third ($0.02 \leq \xi \leq 0.22$) and small fourth ($0.22 \leq \xi \leq 0.4$) peaks occupy the Perseus arm region \citep{Dambis2015}. In turn, the depressions at $\xi \approx -0.15$ and $\xi \approx 0.03$ approximately coincide with the interarm intervals. Such an effect can be explained by the presence in the arms of associations covering a larger range of ages, i.e. not only young, but also the older ones. 

\begin{figure}
	\includegraphics[width=\columnwidth]{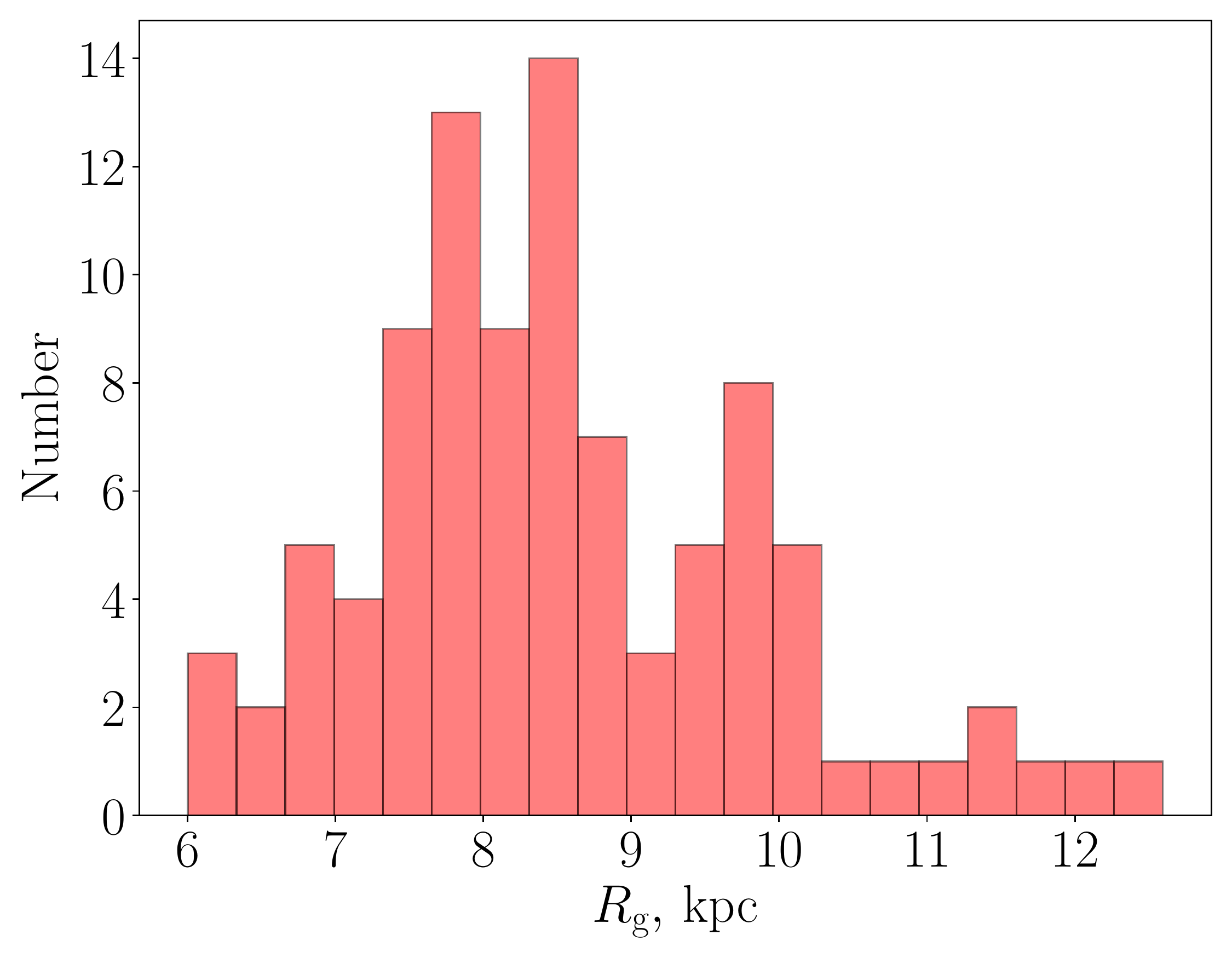}
	\caption{Distribution of clusters with $T_{\mathrm l} \leq 200$ Myr and $t_{\mathrm{obs,l}} \geq t_{\mathrm{theor}}(1 \sigma)$ as a function of Galactocentric distance, $R_{\mathrm g}$.}
	\label{fig:hist_rg}
\end{figure}

\begin{figure}
	\includegraphics[width=\columnwidth]{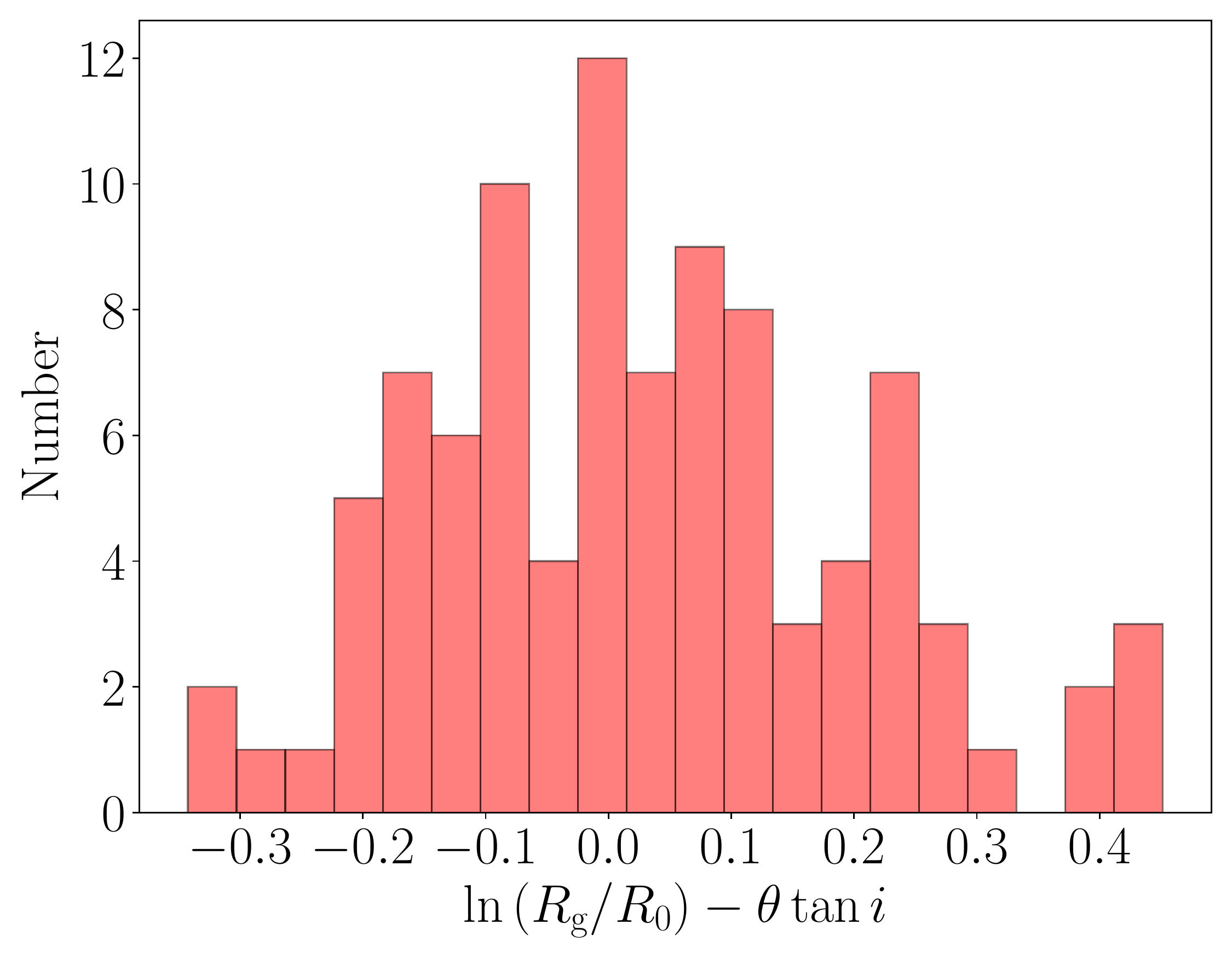}
	\caption{As Figure~\ref{fig:hist_rg}, but for $\xi = \ln{(R_{\mathrm g} / R_0)} - \theta \tan(i)$ instead of Galactocentric distance, $R_{\mathrm g}$.}
	\label{fig:hist_rspir}
\end{figure}

We also show the distribution of the clusters examined as a function of Galactocentric distance and of the $\xi$ parameter, in Figures~\ref{fig:hist_rg} and \ref{fig:hist_rspir}, respectively. The sample of final clusters covers a Galactocentric distance range of more than 6 kpc, and in their distribution in Figures~\ref{fig:hist_rspir} one can also trace an increased number of clusters in the region of the Car--Sag and Perseus arms, but much less clearly, owing, in part, to a decrease in the number of clusters with distance from the Sun, the position of which corresponds to the value $\xi = 0$.

Finally, among the members of some of the clusters for which age estimates were obtained, some previously known OCs were included. We compared the ages of these OCs \citep{Dias2021} with our newly derived $T_{\mathrm l}$ ages for the respective clusters: see Table~\ref{tab:ocs_in_assocs}. We found 32 OCs whose chronological ages are consistent with the kinematic age of the parent cluster (among clusters showing signs of general expansion along Galactic longitude at the 1$\sigma$ confidence level), considering the prevailing uncertainties. This agreement may serve as evidence of a real connection between the clusters studied and the OCs, coupled with their possible origin in a single process of hierarchical star formation.
\begin{table*}
	\centering
	\caption{Comparison of the ages of the newly found clusters (which show signs of general expansion at the 1$\sigma$ confidence level) with the ages of the OCs contained within them. $T_{\mathrm{OC}}$, $T_{\mathrm{OC, low}}$, $T_{\mathrm{OC, high}}$: average, minimum and maximum estimates, respectively, of the age of the OC from the \citet{Dias2021} catalogue; $T_{\mathrm{clus}}$, $T_{\mathrm{clus, low}}$, $T_{\mathrm{clus, high}}$: kinematic cluster ages obtained in this paper. Cl: label of the parent cluster.}
	\label{tab:ocs_in_assocs}
	\begin{tabular}{lccccccc} 
		\hline
		OC & $T_{\mathrm{OC}}$, & $T_{\mathrm{OC, low}}$, & $T_{\mathrm{OC, high}}$, & Cl & $T_{\mathrm{clus}}$, & $T_{\mathrm{clus, low}}$, & $T_{\mathrm{clus, high}}$,\\
		& Myr & Myr & Myr & & Myr & Myr & Myr\\
		\hline
           Alessi\_44 & 113.76 & 73.28 & 176.60 & 49 & 116.52 & 106.80 &  126.24\\
              BH\_121 &   7.31 &  5.77 &   9.27 & 205 & 7.12 & 6.92 & 7.31\\
          ESO\_368\_14 & 172.19 & 31.41 & 944.06 & 68 & 82.95 &  76.82 &   89.08\\
             Hogg\_19 &  13.40 &  5.26 &  34.12 & 157 & 14.24 &  14.07 &   14.41\\
             IC\_2581 &  12.85 &  7.66 &  21.58 & 212 & 19.92 &  19.24 &   20.61\\
             IC\_2948 &   7.41 &  3.12 &  17.62& 205 & 7.12 &   6.92 &    7.31\\
             IC\_4996 &  11.69 &  9.44 &  14.49 & 79 & 11.35 &  11.16 &   11.54\\
             LP\_0733 &  48.87 & 28.77 &  82.99 & 25 & 40.09 &  38.94 &   41.23\\
             LP\_1429 & 175.39 & 81.66 & 376.70 & 70 & 150.93 & 127.26 &  174.60\\
             LP\_1728 &  11.30 &  2.24 &  57.02 & 15 & 42.60 &  40.80 &   44.41\\
            NGC\_2362 &  14.32 & 10.62 &  19.32 & 64 & 12.55 &  12.28 &   12.82\\
            NGC\_2414 &  56.49 & 32.21 &  99.08 & 17 & 70.53 &  64.48 &   76.59\\
            NGC\_5662 & 105.44 & 68.87 & 161.44 & 170 & 64.91 &  58.42 &   71.40\\
            NGC\_6087 & 100.69 & 86.50 & 117.22 & 182 & 96.37 &  87.18 &  105.57\\
            NGC\_6200 &  13.74 & 11.97 &  15.78 & 157 & 14.24 &  14.07 &   14.41\\
            NGC\_6531 &  12.76 &  9.98 &  16.33 & 200 & 16.92 &  16.03 &   17.82\\
             NGC\_869 &  12.88 & 11.72 &  14.16 & 143 & 13.21 &  13.14 &   13.29\\
           Roslund\_2 &  12.02 & 10.79 &  13.40 & 75 & 13.97 &  13.26 &   14.68\\
        Ruprecht\_119 & 101.16 & 53.09 & 192.75 & 158 & 78.82 &  69.04 &   88.60\\
        Ruprecht\_148 &  79.98 & 29.11 & 219.79 & 58 & 42.60 &  42.00 &   43.21\\
         Ruprecht\_35 & 153.46 & 46.34 & 508.16 & 68 & 82.95 &  76.82 &   89.08\\
         Ruprecht\_47 & 139.32 & 88.31 & 219.79 & 70 & 150.93 & 127.26 &  174.60\\
         Ruprecht\_65 &  45.29 & 19.95 & 102.80 & 107 & 68.09 &  62.42 &   73.75\\
             UBC\_191 &  15.31 &  7.35 &  31.92 & 143 & 13.21 &  13.14 &   13.29\\
             UBC\_192 &  18.97 & 12.65 &  28.44 & 143 & 13.21 &  13.14 &   13.29\\
              UBC\_31 &  26.18 &  8.38 &  81.85 & 35 & 19.93 &  19.06 &   20.80\\
             UBC\_437 &  80.17 & 42.66 & 150.66 & 114 & 93.79 &  85.20 &  102.38\\
             UBC\_483 &  32.73 &  9.40 & 114.02 & 107 & 68.09 &  62.42 &   73.75\\
             UBC\_606 &  10.00&  5.81 &  17.22 & 143 & 13.21 &  13.14 &   13.29\\
             UBC\_664 &  11.43 &  3.27 &  39.99 & 47 & 23.39 &  22.91 &   23.87\\
              UPK\_45 &  95.50 & 51.05 & 178.65 & 49 & 116.52 & 106.80 &  126.24\\
             UPK\_630 & 112.72 & 88.31 & 143.88 & 185 & 89.03 &  83.13 &   94.92\\
		\hline
	\end{tabular}
\end{table*}

\subsection{Comparison with historical associations}\label{historical}
\begin{figure*}
	\includegraphics[width=\textwidth]{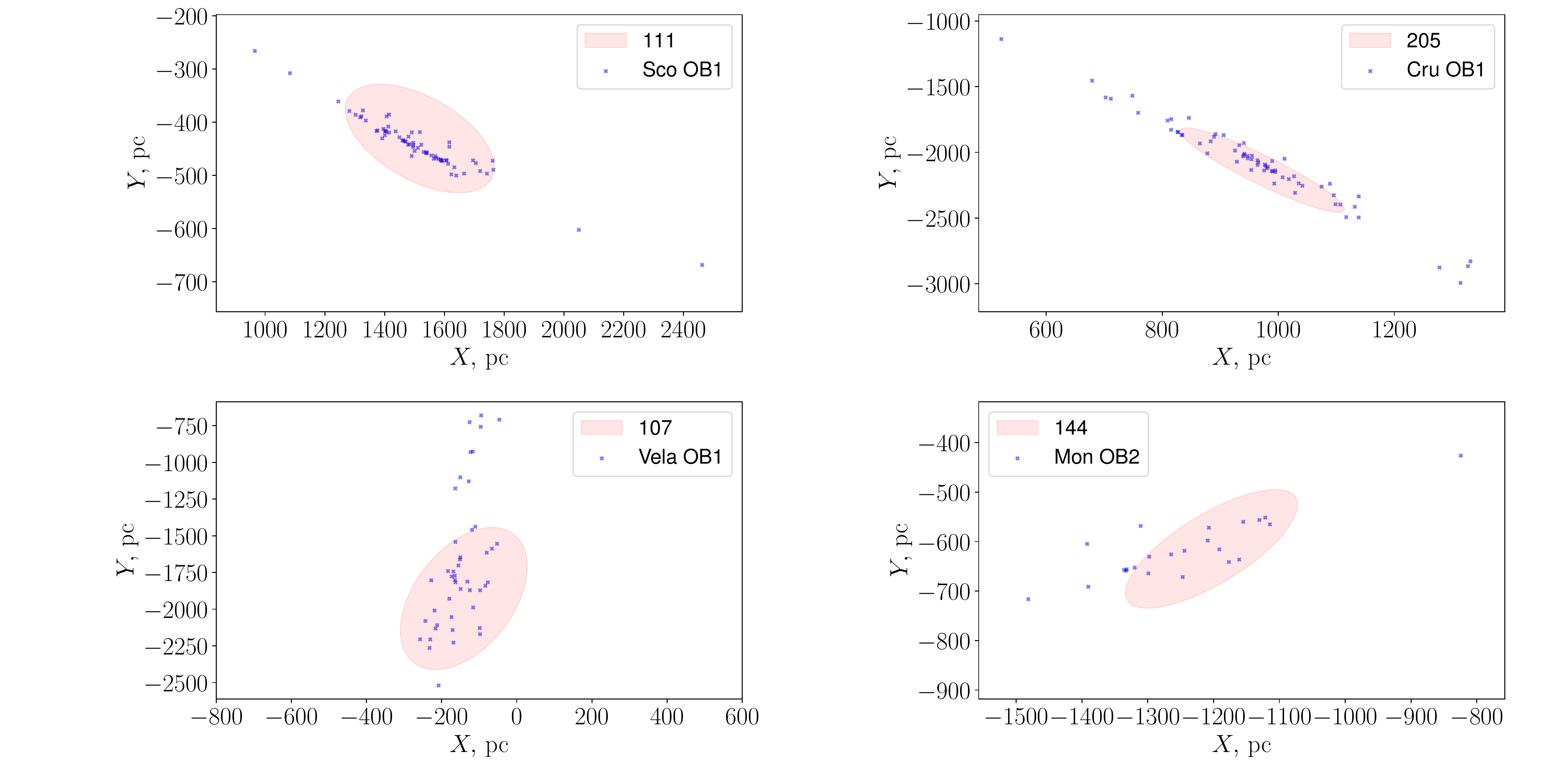}
	\caption{Correspondences between the positions of the clusters and associations Sco OB1, Cru OB1, Vela OB1 and Mon OB2. Filled pale-red ellipses represent the fitting of cluster members with a two-dimensional Gaussian distribution at the 3$\sigma$ level. Blue crosses are association members from \citet{MD20}.}
	\label{fig:assoc_examples}
\end{figure*}

We performed a comparison with the associations studied by \citet{MD17} and \citet{MD20}. These studies were based on the partitioning of young stars into associations by \citet{BlahaHump1989}. The heliocentric distances to the latter were refined using data from \citet{BJ2021} to convert all distances to the {\sl Gaia} EDR3 system. In many cases, the clusters we obtained reproduce the position of the associations we investigated in the Galaxy, including large and widely studied associations such as Sco--Cen (cluster No. 45), Perseus (Per) OB1 (cluster No. 143), Per OB2 (cluster No. 35), Orion (Ori) OB1 (clusters Nos 28, 29), Gemini (Gem) OB1 (cluster No. 114) and others. However, note that, first, the use of refined \citet{BJ2021} distances removes many stars from the \citet{BlahaHump1989} division into excessively large or small distances, owing to which they cease to be members of the main concentration of young stars in historical associations; second, in a number of cases there is a cross-correspondence, when several of our clusters represent parts of one historical association or, vice versa, several associations turn out to be members of one cluster. For example, associations Cassiopeia (Cas) OB2 and Cas OB5 are parts of one cluster, No. 141; Cep OB1 includes clusters Nos 80, 120, 121 and 122, whereas Sgr OB1 includes clusters Nos 199 and 200. This makes it more difficult to manually identify a direct correspondence between historical associations and our newly identified clusters. However, at the same time, the presence of such cross-identifications may indicate the presence of substructure and aggregates of associations, which have a common origin, similar to what has been shown for OCs by \citet{Paunzen2021}. Figure~\ref{fig:assoc_examples} shows examples of correspondences between the positions of clusters and the associations Sco OB1, Crux (Cru) OB1, Vela OB1 and Monoceros (Mon) OB2 from \citet{MD20}, where, for greater clarity, the clusters we found are indicated by filled ellipses, which represent the fitting of cluster members with a two-dimensional Gaussian distribution at the 3$\sigma$ level.

In addition, even in cases where the correspondence between the cluster and the association is unambiguous, the cluster composition does not usually exactly reproduce the composition of the corresponding historical association. A similar scenario, where the redefinition of associations using modern methods reveals a discrepancy compared with their historical composition, is described by \citet{CYGassocs2021}, where the authors investigated the OB associations in Cygnus. Thus, once again we draw attention to the fact that the historical composition of associations may not correspond to physical reality and must hence be revised. 

\section{Conclusions}

We compiled a well-defined sample of OB stars based on the LAMOST DR5 \citep{LamostDR5}, \citet{Xu2018} and \citet{Skiff} catalogues, supplemented with young OCs from \citet{Dias2021}. The resulting sample contains more than 47,700 young objects located in the Galactic thin disc. 

To search for candidates for OB associations, we applied the HDBSCAN$^{*}$ clustering algorithm in the five-dimensional space of the Cartesian heliocentric coordinates $X,\,Y,\,Z$ and proper motions $PM_{\mathrm \alpha},\,PM_{\mathrm \delta}$. We first corrected these parameters to unit variance to facilitate the algorithm's operation. For \textit{min\_cluster\_size} = 10 in the HDBSCAN$^{*}$ algorithm, we obtained a total of 214 clusters with a median diameter $D_{\mathrm{med}} = 98.7 \mathrm{\,pc}$, a median number of members $N_{\mathrm{med}} = 20$ and a median one-dimensional internal velocity dispersion $\sigma V_{\mathrm{med}} = 2.36 \mathrm{\,km\,s}^{-1}$. We estimated the cluster size as the diameter of the circle in the plane of the sky containing 68 per cent of an association's members. To determine the true one-dimensional internal velocity dispersion of the candidate associations thus identified, we corrected the observed values for two main effects, including errors in the observed proper motions and the radial velocities of the clusters. The latter may result in a visible contraction or expansion. Because of the lack of {\it bona fide} observational estimates of radial velocities for bright OB stars we were able to use the observed radial velocities only for a small number of clusters. For those clusters where such estimates were unavailable, we simulated radial velocities using a numerical model, assuming that associations move in circular orbits about the Galactic centre. This assumption is justified since molecular clouds, as components of a dissipative subsystem, should move in circular orbits, and the associations forming from them inherit this motion, without having had time to undergo significant dynamical heating during their lifetimes of just several tens of Myr. The sizes of the resulting clusters and their internal velocity dispersions are in good agreement with those expected for OB associations.

We investigated the proper motion--coordinate diagrams along Galactic longitude and latitude for all clusters using the Monte Carlo method. For each cluster, 90 per cent of random members were selected 100 times, and kinematic age estimates were obtained from the slope of the approximating line in the diagram. The final ages were defined as the medians and median absolute deviations of the series of values thus obtained. To assess the reliability of the correlations between proper motions and the corresponding coordinates (i.e., the reliability of our determination of general expansion signatures), we used the Student's $t$ test, comparing the observed values of the $t$ parameter with the pre-calculated theoretical values for confidence levels corresponding to 1$\sigma$, 2$\sigma$ and 3$\sigma$ of a Gaussian distribution. Among clusters younger than 200 Myr, 95 clusters show a correlation at the 1$\sigma$ level of confidence in longitude and 89 in latitude (47 in both), 38 clusters show a correlation at the 3$\sigma$ level of confidence in longitude and 25 in latitude (11 in both). The characteristic ages of these clusters are several tens of Myr, which is as expected for typical OB associations in the Galaxy.

The dependence of the mean age on Galactocentric distance and on the parameter $\xi = \ln{(R_{\mathrm g} / R_0)} - \theta \tan(i)$ demonstrated a correlation between the positions of peaks and depressions with Galactic spiral arms. The positions of local peaks, marking areas of increased mean age, correlate with the position of the Inner, Car--Sgr and Perseus arms. Such an effect can be explained by the presence in the arms of associations spanning a larger range of ages, i.e., not only young but also older associations. With an increase in the range of galactocentric distances covered by future surveys and an increase in the statistics of the clusters studied in future sky surveys, we hope for additional verification and refinement of this effect and, possibly, the detection of an outer arm as well.

We also compared the ages of OCs from \citet{Dias2021} with the ages of the resulting clusters hosting them. We found 32 OCs with ages consistent with the kinematic ages of the parent clusters (among clusters showing signs of general expansion along Galactic longitude at the 1$\sigma$ confidence level), considering the prevailing uncertainties. This agreement offers evidence of a real connection between the clusters studied and the OCs, coupled with their possible origin in a single process of hierarchical star formation. 

We manually compared the clusters identified in this paper with the associations from \citet{MD17} and \citet{MD20}, taking into account the updated heliocentric distances of the latter based on {\sl Gaia} EDR3 \citep{BJ2021}. In many cases, the clusters we obtained reproduce the position of the investigated associations in the Galaxy, including large and widely studied associations. However, in a number of cases there is a cross-correspondence, where several of our clusters represent parts of one historical association or, vice versa, several associations turn out to be members of one cluster. This makes it more difficult to manually identify a direct correspondence between historical associations and our clusters, but at the same time, the presence of such cross-identifications may indicate the presence of substructure and aggregates of associations, which have a common origin, similar to what has been shown for OCs by \citet{Paunzen2021}. Thus, once again we suggest that the historical compositions of associations may not correspond to reality and must be revised. Analysis of such scenarios and a more detailed study of individual clusters will also be one of the goals of our future work.

\section*{Acknowledgements}

We thank A.S. Rastorguev for his help and valuable advice. AAC, EVG and AKD are grateful to the Russian Foundation for Basic Research for partial financial support (project No. 19-02-00611). This research was also supported in part by the Australian Research Council Centre of Excellence for All Sky Astrophysics in 3 Dimensions (ASTRO 3D) through project No. CE170100013. This work has made use of data from the European Space Agency's (ESA) space mission {\sl Gaia} (\url{https://www.cosmos.esa.int/gaia}), processed by the {\sl Gaia} Data Processing and Analysis Consortium (DPAC, \url{https://www.cosmos.esa.int/web/gaia/dpac/consortium}). Funding for the DPAC is provided by national institutions, in particular the institutions participating in the {\sl Gaia} MultiLateral Agreement (MLA). We also made use of the VizieR catalogue access tool, CDS, Strasbourg, France (\url{https://vizier.u-strasbg.fr/viz-bin/VizieR}). The original description of the VizieR service was published in \citet{Vizier2000}.

\section*{Data Availability}

Information about the parameters of the resulting clusters (including estimates of kinematic ages), as well as list of their members is available as supplementary material.



\bibliographystyle{mnras}
\bibliography{example} 

\begin{thebibliography}{}
\makeatletter
\relax
\def\mn@urlcharsother{\let\do\@makeother \do\$\do\&\do\#\do\^\do\_\do\%\do\~}
\def\mn@doi{\begingroup\mn@urlcharsother \@ifnextchar [ {\mn@doi@}
  {\mn@doi@[]}}
\def\mn@doi@[#1]#2{\def\@tempa{#1}\ifx\@tempa\@empty \href
  {http://dx.doi.org/#2} {doi:#2}\else \href {http://dx.doi.org/#2} {#1}\fi
  \endgroup}
\def\mn@eprint#1#2{\mn@eprint@#1:#2::\@nil}
\def\mn@eprint@arXiv#1{\href {http://arxiv.org/abs/#1} {{\tt arXiv:#1}}}
\def\mn@eprint@dblp#1{\href {http://dblp.uni-trier.de/rec/bibtex/#1.xml}
  {dblp:#1}}
\def\mn@eprint@#1:#2:#3:#4\@nil{\def\@tempa {#1}\def\@tempb {#2}\def\@tempc
  {#3}\ifx \@tempc \@empty \let \@tempc \@tempb \let \@tempb \@tempa \fi \ifx
  \@tempb \@empty \def\@tempb {arXiv}\fi \@ifundefined
  {mn@eprint@\@tempb}{\@tempb:\@tempc}{\expandafter \expandafter \csname
  mn@eprint@\@tempb\endcsname \expandafter{\@tempc}}}

\bibitem[\protect\citeauthoryear{{Allen} et~al.,}{{Allen}
  et~al.}{2007}]{Allen2007}
{Allen} L.,  et~al., 2007, in {Reipurth} B.,  {Jewitt} D.,   {Keil} K.,  eds,
  Protostars and Planets V. p.~361 (\mn@eprint {arXiv} {astro-ph/0603096})

\bibitem[\protect\citeauthoryear{{Ambartsumian}}{{Ambartsumian}}{1947}]{Amb1947}
{Ambartsumian} V.~A.,  1947, {The evolution of stars and astrophysics}

\bibitem[\protect\citeauthoryear{{Bailer-Jones}, {Rybizki}, {Fouesneau},
  {Demleitner}  \& {Andrae}}{{Bailer-Jones} et~al.}{2021}]{BJ2021}
{Bailer-Jones} C.~A.~L.,  {Rybizki} J.,  {Fouesneau} M.,  {Demleitner} M.,
  {Andrae} R.,  2021, \mn@doi [\aj] {10.3847/1538-3881/abd806}, \href
  {https://ui.adsabs.harvard.edu/abs/2021AJ....161..147B} {161, 147}

\bibitem[\protect\citeauthoryear{{Baumgardt} \& {Kroupa}}{{Baumgardt} \&
  {Kroupa}}{2007}]{Baumgardt2007}
{Baumgardt} H.,  {Kroupa} P.,  2007, \mn@doi [\mnras]
  {10.1111/j.1365-2966.2007.12209.x}, \href
  {https://ui.adsabs.harvard.edu/abs/2007MNRAS.380.1589B} {380, 1589}

\bibitem[\protect\citeauthoryear{{Becker}}{{Becker}}{1963}]{Becker1963}
{Becker} W.,  1963, \zap, \href
  {https://ui.adsabs.harvard.edu/abs/1963ZA.....57..117B} {57, 117}

\bibitem[\protect\citeauthoryear{Binney \& Tremaine}{Binney \&
  Tremaine}{2011}]{BinTre2nd}
Binney J.,  Tremaine S.,  2011, Galactic Dynamics: Second Edition.
Princeton Series in Astrophysics, Princeton University Press

\bibitem[\protect\citeauthoryear{{Blaauw}}{{Blaauw}}{1952a}]{Blaauw1952a}
{Blaauw} A.,  1952a, \bain, \href
  {https://ui.adsabs.harvard.edu/abs/1952BAN....11..405B} {11, 405}

\bibitem[\protect\citeauthoryear{{Blaauw}}{{Blaauw}}{1952b}]{Blaauw1952b}
{Blaauw} A.,  1952b, \bain, \href
  {https://ui.adsabs.harvard.edu/abs/1952BAN....11..414B} {11, 414}

\bibitem[\protect\citeauthoryear{{Blaauw}}{{Blaauw}}{1964}]{Blaauw1964}
{Blaauw} A.,  1964, \mn@doi [\araa] {10.1146/annurev.aa.02.090164.001241},
  \href {https://ui.adsabs.harvard.edu/abs/1964ARA&A...2..213B} {2, 213}

\bibitem[\protect\citeauthoryear{{Blaha} \& {Humphreys}}{{Blaha} \&
  {Humphreys}}{1989}]{BlahaHump1989}
{Blaha} C.,  {Humphreys} R.~M.,  1989, \mn@doi [\aj] {10.1086/115244}, \href
  {https://ui.adsabs.harvard.edu/abs/1989AJ.....98.1598B} {98, 1598}

\bibitem[\protect\citeauthoryear{{Bobylev} \& {Bajkova}}{{Bobylev} \&
  {Bajkova}}{2019}]{BoBa2019}
{Bobylev} V.~V.,  {Bajkova} A.~T.,  2019, \mn@doi [Astronomy Letters]
  {10.1134/S106377371906001X}, \href
  {https://ui.adsabs.harvard.edu/abs/2019AstL...45..331B} {45, 331}

\bibitem[\protect\citeauthoryear{{Brown}, {Dekker}  \& {de Zeeuw}}{{Brown}
  et~al.}{1997}]{Brown1997}
{Brown} A.~G.~A.,  {Dekker} G.,   {de Zeeuw} P.~T.,  1997, \mn@doi [\mnras]
  {10.1093/mnras/285.3.479}, \href
  {https://ui.adsabs.harvard.edu/abs/1997MNRAS.285..479B} {285, 479}

\bibitem[\protect\citeauthoryear{{Cantat-Gaudin} et~al.,}{{Cantat-Gaudin}
  et~al.}{2018}]{CG2018}
{Cantat-Gaudin} T.,  et~al., 2018, \mn@doi [\aap]
  {10.1051/0004-6361/201833476}, \href
  {https://ui.adsabs.harvard.edu/abs/2018A&A...618A..93C} {618, A93}

\bibitem[\protect\citeauthoryear{{Dambis} et~al.,}{{Dambis}
  et~al.}{2015}]{Dambis2015}
{Dambis} A.~K.,  et~al., 2015, \mn@doi [Astronomy Letters]
  {10.1134/S1063773715090017}, \href
  {https://ui.adsabs.harvard.edu/abs/2015AstL...41..489D} {41, 489}

\bibitem[\protect\citeauthoryear{{Deng} et~al.,}{{Deng}
  et~al.}{2012}]{Deng2012}
{Deng} L.-C.,  et~al., 2012, \mn@doi [Research in Astronomy and Astrophysics]
  {10.1088/1674-4527/12/7/003}, \href
  {https://ui.adsabs.harvard.edu/abs/2012RAA....12..735D} {12, 735}

\bibitem[\protect\citeauthoryear{{Dias}, {Monteiro}, {Moitinho}, {L{\'e}pine},
  {Carraro}, {Paunzen}, {Alessi}  \& {Villela}}{{Dias} et~al.}{2021}]{Dias2021}
{Dias} W.~S.,  {Monteiro} H.,  {Moitinho} A.,  {L{\'e}pine} J.~R.~D.,
  {Carraro} G.,  {Paunzen} E.,  {Alessi} B.,   {Villela} L.,  2021, \mn@doi
  [\mnras] {10.1093/mnras/stab770}, \href
  {https://ui.adsabs.harvard.edu/abs/2021MNRAS.504..356D} {504, 356}

\bibitem[\protect\citeauthoryear{Ester, Kriegel, Sander  \& Xu}{Ester
  et~al.}{1996}]{Ester96adensity-based}
Ester M.,  Kriegel H.-P.,  Sander J.,   Xu X.,  1996. AAAI Press, pp 226--231

\bibitem[\protect\citeauthoryear{{Evans} Neal~J. et~al.,}{{Evans}
  et~al.}{2009}]{Evans2009}
{Evans} Neal~J. I.,  et~al., 2009, \mn@doi [\apjs]
  {10.1088/0067-0049/181/2/321}, \href
  {https://ui.adsabs.harvard.edu/abs/2009ApJS..181..321E} {181, 321}

\bibitem[\protect\citeauthoryear{{Flynn}, {Holmberg}, {Portinari}, {Fuchs}  \&
  {Jahrei{\ss}}}{{Flynn} et~al.}{2006}]{Flynn2006}
{Flynn} C.,  {Holmberg} J.,  {Portinari} L.,  {Fuchs} B.,   {Jahrei{\ss}} H.,
  2006, \mn@doi [\mnras] {10.1111/j.1365-2966.2006.10911.x}, \href
  {https://ui.adsabs.harvard.edu/abs/2006MNRAS.372.1149F} {372, 1149}

\bibitem[\protect\citeauthoryear{{Gaia Collaboration} et~al.,}{{Gaia
  Collaboration} et~al.}{2016}]{GaiaMission2016}
{Gaia Collaboration} et~al., 2016, \mn@doi [\aap]
  {10.1051/0004-6361/201629272}, \href
  {https://ui.adsabs.harvard.edu/abs/2016A&A...595A...1G} {595, A1}

\bibitem[\protect\citeauthoryear{{Gaia Collaboration} et~al.,}{{Gaia
  Collaboration} et~al.}{2018}]{GaiaDR2_2018}
{Gaia Collaboration} et~al., 2018, \mn@doi [\aap]
  {10.1051/0004-6361/201833051}, \href
  {https://ui.adsabs.harvard.edu/abs/2018A&A...616A...1G} {616, A1}

\bibitem[\protect\citeauthoryear{{Gaia Collaboration} et~al.,}{{Gaia
  Collaboration} et~al.}{2021}]{GaiaEDR3_2021}
{Gaia Collaboration} et~al., 2021, \mn@doi [\aap]
  {10.1051/0004-6361/202039657}, \href
  {https://ui.adsabs.harvard.edu/abs/2021A&A...649A...1G} {649, A1}

\bibitem[\protect\citeauthoryear{{Garmany} \& {Stencel}}{{Garmany} \&
  {Stencel}}{1992}]{Garmany1992}
{Garmany} C.~D.,  {Stencel} R.~E.,  1992, \aaps, \href
  {https://ui.adsabs.harvard.edu/abs/1992A&AS...94..211G} {94, 211}

\bibitem[\protect\citeauthoryear{{Goodwin} \& {Bastian}}{{Goodwin} \&
  {Bastian}}{2006}]{Goodwin2006}
{Goodwin} S.~P.,  {Bastian} N.,  2006, \mn@doi [\mnras]
  {10.1111/j.1365-2966.2006.11078.x}, \href
  {https://ui.adsabs.harvard.edu/abs/2006MNRAS.373..752G} {373, 752}

\bibitem[\protect\citeauthoryear{{Hills}}{{Hills}}{1980}]{Hills1980}
{Hills} J.~G.,  1980, \mn@doi [\apj] {10.1086/157703}, \href
  {https://ui.adsabs.harvard.edu/abs/1980ApJ...235..986H} {235, 986}

\bibitem[\protect\citeauthoryear{{Humphreys}}{{Humphreys}}{1978}]{Humphreys1978}
{Humphreys} R.~M.,  1978, \mn@doi [\apjs] {10.1086/190559}, \href
  {https://ui.adsabs.harvard.edu/abs/1978ApJS...38..309H} {38, 309}

\bibitem[\protect\citeauthoryear{{Kroupa}, {Aarseth}  \& {Hurley}}{{Kroupa}
  et~al.}{2001}]{Kroupa2001}
{Kroupa} P.,  {Aarseth} S.,   {Hurley} J.,  2001, \mn@doi [\mnras]
  {10.1046/j.1365-8711.2001.04050.x}, \href
  {https://ui.adsabs.harvard.edu/abs/2001MNRAS.321..699K} {321, 699}

\bibitem[\protect\citeauthoryear{{Lada} \& {Lada}}{{Lada} \&
  {Lada}}{2003}]{Lada2003}
{Lada} C.~J.,  {Lada} E.~A.,  2003, \mn@doi [\araa]
  {10.1146/annurev.astro.41.011802.094844}, \href
  {https://ui.adsabs.harvard.edu/abs/2003ARA&A..41...57L} {41, 57}

\bibitem[\protect\citeauthoryear{{Lamb}, {Oey}, {Werk}  \& {Ingleby}}{{Lamb}
  et~al.}{2010}]{Lamb2010}
{Lamb} J.~B.,  {Oey} M.~S.,  {Werk} J.~K.,   {Ingleby} L.~D.,  2010, \mn@doi
  [\apj] {10.1088/0004-637X/725/2/1886}, \href
  {https://ui.adsabs.harvard.edu/abs/2010ApJ...725.1886L} {725, 1886}

\bibitem[\protect\citeauthoryear{{Lim}, {Naz{\'e}}, {Gosset}  \& {Rauw}}{{Lim}
  et~al.}{2019}]{Lim2019}
{Lim} B.,  {Naz{\'e}} Y.,  {Gosset} E.,   {Rauw} G.,  2019, \mn@doi [\mnras]
  {10.1093/mnras/stz2548}, \href
  {https://ui.adsabs.harvard.edu/abs/2019MNRAS.490..440L} {490, 440}

\bibitem[\protect\citeauthoryear{{Luo}, {Zhao}, {Zhao}  \& {et al.}}{{Luo}
  et~al.}{2019}]{LamostDR5}
{Luo} A.~L.,  {Zhao} Y.~H.,  {Zhao} G.,   {et al.} 2019, VizieR Online Data
  Catalog, \href {https://ui.adsabs.harvard.edu/abs/2019yCat.5164....0L} {p.
  V/164}

\bibitem[\protect\citeauthoryear{{Markarian}}{{Markarian}}{1951}]{Markarian1951}
{Markarian} B.~E.,  1951, Communications of the Byurakan Astrophysical
  Observatory, \href {https://ui.adsabs.harvard.edu/abs/1951CoBAO...9....3M}
  {9, 3}

\bibitem[\protect\citeauthoryear{McInnes, Healy  \& Astels}{McInnes
  et~al.}{2017}]{hdbscan2017}
McInnes L.,  Healy J.,   Astels S.,  2017, \mn@doi [The Journal of Open Source
  Software] {10.21105/joss.00205}, 2

\bibitem[\protect\citeauthoryear{{Mel'Nik} \& {Dambis}}{{Mel'Nik} \&
  {Dambis}}{2009}]{MelDam2009}
{Mel'Nik} A.~M.,  {Dambis} A.~K.,  2009, \mn@doi [\mnras]
  {10.1111/j.1365-2966.2009.15484.x}, \href
  {https://ui.adsabs.harvard.edu/abs/2009MNRAS.400..518M} {400, 518}

\bibitem[\protect\citeauthoryear{{Mel'Nik} \& {Efremov}}{{Mel'Nik} \&
  {Efremov}}{1995}]{ME1995}
{Mel'Nik} A.~M.,  {Efremov} Y.~N.,  1995, Astronomy Letters, \href
  {https://ui.adsabs.harvard.edu/abs/1995AstL...21...10M} {21, 10}

\bibitem[\protect\citeauthoryear{{Mel'nik} \& {Dambis}}{{Mel'nik} \&
  {Dambis}}{2017}]{MD17}
{Mel'nik} A.~M.,  {Dambis} A.~K.,  2017, \mn@doi [\mnras]
  {10.1093/mnras/stx2225}, \href
  {https://ui.adsabs.harvard.edu/abs/2017MNRAS.472.3887M} {472, 3887}

\bibitem[\protect\citeauthoryear{{Melnik} \& {Dambis}}{{Melnik} \&
  {Dambis}}{2020}]{MD20}
{Melnik} A.~M.,  {Dambis} A.~K.,  2020, \mn@doi [\mnras]
  {10.1093/mnras/staa454}, \href
  {https://ui.adsabs.harvard.edu/abs/2020MNRAS.493.2339M} {493, 2339}

\bibitem[\protect\citeauthoryear{{Morgan}, {Sharpless}  \&
  {Osterbrock}}{{Morgan} et~al.}{1952}]{Morgan1952}
{Morgan} W.~W.,  {Sharpless} S.,   {Osterbrock} D.,  1952, \mn@doi [\aj]
  {10.1086/106673}, \href
  {https://ui.adsabs.harvard.edu/abs/1952AJ.....57....3M} {57, 3}

\bibitem[\protect\citeauthoryear{{Morgan}, {Whitford}  \& {Code}}{{Morgan}
  et~al.}{1953}]{Morgan1953}
{Morgan} W.~W.,  {Whitford} A.~E.,   {Code} A.~D.,  1953, \mn@doi [\apj]
  {10.1086/145754}, \href
  {https://ui.adsabs.harvard.edu/abs/1953ApJ...118..318M} {118, 318}

\bibitem[\protect\citeauthoryear{{Ochsenbein}, {Bauer}  \&
  {Marcout}}{{Ochsenbein} et~al.}{2000}]{Vizier2000}
{Ochsenbein} F.,  {Bauer} P.,   {Marcout} J.,  2000, \mn@doi [\aaps]
  {10.1051/aas:2000169}, \href
  {https://ui.adsabs.harvard.edu/abs/2000A&AS..143...23O} {143, 23}

\bibitem[\protect\citeauthoryear{{Penoyre}, {Belokurov}  \& {Evans}}{{Penoyre}
  et~al.}{2021}]{Penoyre2021}
{Penoyre} Z.,  {Belokurov} V.,   {Evans} N.~W.,  2021, arXiv e-prints, \href
  {https://ui.adsabs.harvard.edu/abs/2021arXiv211110380P} {p. arXiv:2111.10380}

\bibitem[\protect\citeauthoryear{{Piecka} \& {Paunzen}}{{Piecka} \&
  {Paunzen}}{2021}]{Paunzen2021}
{Piecka} M.,  {Paunzen} E.,  2021, arXiv e-prints, \href
  {https://ui.adsabs.harvard.edu/abs/2021arXiv210608920P} {p. arXiv:2106.08920}

\bibitem[\protect\citeauthoryear{{Quintana} \& {Wright}}{{Quintana} \&
  {Wright}}{2021}]{CYGassocs2021}
{Quintana} A.~L.,  {Wright} N.~J.,  2021, \mn@doi [\mnras]
  {10.1093/mnras/stab2663}, \href
  {https://ui.adsabs.harvard.edu/abs/2021MNRAS.508.2370Q} {508, 2370}

\bibitem[\protect\citeauthoryear{{Rastorguev}, {Utkin}, {Zabolotskikh},
  {Dambis}, {Bajkova}  \& {Bobylev}}{{Rastorguev} et~al.}{2017}]{RastUt17}
{Rastorguev} A.~S.,  {Utkin} N.~D.,  {Zabolotskikh} M.~V.,  {Dambis} A.~K.,
  {Bajkova} A.~T.,   {Bobylev} V.~V.,  2017, \mn@doi [Astrophysical Bulletin]
  {10.1134/S1990341317020043}, \href
  {https://ui.adsabs.harvard.edu/abs/2017AstBu..72..122R} {72, 122}

\bibitem[\protect\citeauthoryear{{Skiff}}{{Skiff}}{2014}]{Skiff}
{Skiff} B.~A.,  2014, VizieR Online Data Catalog, \href
  {https://ui.adsabs.harvard.edu/abs/2014yCat....1.2023S} {p.~B/mk}

\bibitem[\protect\citeauthoryear{{Vaher}}{{Vaher}}{2020}]{Vaher2020}
{Vaher} E.,  2020, \mn@doi [Research Notes of the American Astronomical
  Society] {10.3847/2515-5172/aba952}, \href
  {https://ui.adsabs.harvard.edu/abs/2020RNAAS...4..116V} {4, 116}

\bibitem[\protect\citeauthoryear{{Ward} \& {Kruijssen}}{{Ward} \&
  {Kruijssen}}{2018}]{Ward2018}
{Ward} J.~L.,  {Kruijssen} J.~M.~D.,  2018, \mn@doi [\mnras]
  {10.1093/mnras/sty117}, \href
  {https://ui.adsabs.harvard.edu/abs/2018MNRAS.475.5659W} {475, 5659}

\bibitem[\protect\citeauthoryear{{Ward}, {Kruijssen}  \& {Rix}}{{Ward}
  et~al.}{2020}]{Ward2020}
{Ward} J.~L.,  {Kruijssen} J.~M.~D.,   {Rix} H.-W.,  2020, \mn@doi [\mnras]
  {10.1093/mnras/staa1056}, \href
  {https://ui.adsabs.harvard.edu/abs/2020MNRAS.495..663W} {495, 663}

\bibitem[\protect\citeauthoryear{{Wright} \& {Mamajek}}{{Wright} \&
  {Mamajek}}{2018}]{Wright2018}
{Wright} N.~J.,  {Mamajek} E.~E.,  2018, \mn@doi [\mnras]
  {10.1093/mnras/sty207}, \href
  {https://ui.adsabs.harvard.edu/abs/2018MNRAS.476..381W} {476, 381}

\bibitem[\protect\citeauthoryear{{Xu} et~al.,}{{Xu} et~al.}{2018}]{Xu2018}
{Xu} Y.,  et~al., 2018, \mn@doi [\aap] {10.1051/0004-6361/201833407}, \href
  {https://ui.adsabs.harvard.edu/abs/2018A&A...616L..15X} {616, L15}

\bibitem[\protect\citeauthoryear{{de Grijs} \& {Bono}}{{de Grijs} \&
  {Bono}}{2016}]{deGrijsBono2016}
{de Grijs} R.,  {Bono} G.,  2016, \mn@doi [\apjs] {10.3847/0067-0049/227/1/5},
  \href {https://ui.adsabs.harvard.edu/abs/2016ApJS..227....5D} {227, 5}

\bibitem[\protect\citeauthoryear{{de Zeeuw}, {Hoogerwerf}, {de Bruijne},
  {Brown}  \& {Blaauw}}{{de Zeeuw} et~al.}{1999}]{Zeeuw1999}
{de Zeeuw} P.~T.,  {Hoogerwerf} R.,  {de Bruijne} J.~H.~J.,  {Brown} A.~G.~A.,
   {Blaauw} A.,  1999, \mn@doi [\aj] {10.1086/300682}, \href
  {https://ui.adsabs.harvard.edu/abs/1999AJ....117..354D} {117, 354}

\makeatother
\end{thebibliography}





\bsp	
\label{lastpage}
\end{document}